\documentclass{ar-1col-S2O}
\usepackage[numbers, sort&compress]{natbib}
\usepackage{url}
\jname{Annu. Rev. Condens. Matter Phys.}
\jvol{18}
\jyear{2027}
\doi{10.1146/annurev-conmatphys-XXXXXX-XXXXXX}

\usepackage{times}
\definecolor{rred}{rgb}{0.8, 0.0, 0.0}
\definecolor{bblue}{rgb}{0.0, 0.0, 0.8}
\usepackage[
pagebackref=false,
colorlinks=true,
linkcolor=bblue,
urlcolor=bblue,
filecolor=black,
citecolor=rred,
pdfstartview=FitV,
pdftitle={},
pdfauthor={},
pdfsubject={},
pdfkeywords={},
pdfpagemode=None,
bookmarksopen=true
]{hyperref}

\usepackage{amsmath, amsfonts, amssymb, ascmac, mathtools, bm, braket, comment, tensor, multirow}

\begin{document}

\newcommand{\magenta}[1]{{\textcolor{magenta}{#1}}}
\newcommand{\orange}[1]{{\textcolor{orange}{#1}}}
\newcommand{\teal}[1]{{\textcolor{teal}{#1}}}
\newcommand{\violet}[1]{{\textcolor{violet}{#1}}}

\newcommand{\kk}[1]{{\textcolor{magenta}{#1}}}

\newcommand{\ii}{\text{i}}

\newcommand{\Z}{$\mathbb{Z}$}
\newcommand{\Ztwo}{$\mathbb{Z}_2$}
\newcommand{\twoZ}{$2\mathbb{Z}$}
\newcommand{\ZoplusZ}{$\mathbb{Z} \oplus \mathbb{Z}$}
\newcommand{\ZtwooplusZtwo}{$\mathbb{Z}_2 \oplus \mathbb{Z}_2$}
\newcommand{\twoZoplustwoZ}{$2\mathbb{Z} \oplus 2\mathbb{Z}$}

\newcommand{\Czero}{$\mathrm{U} \left( 2n \right)/\mathrm{U} \left( n \right) \times \mathrm{U} \left( n \right)$}
\newcommand{\Cone}{$\mathrm{U} \left( n \right)$}
\newcommand{\Rzero}{$\mathrm{Sp} \left( 2n \right)/\mathrm{Sp} \left( n \right) \times \mathrm{Sp} \left( n \right)$}
\newcommand{\Rone}{$\mathrm{U} \left( 2n \right)/\mathrm{Sp} \left( n \right)$}
\newcommand{\Rtwo}{$\mathrm{O} \left( 2n \right)/\mathrm{U} \left( n \right)$}
\newcommand{\Rthree}{$\mathrm{O} \left( n \right)$}
\newcommand{\Rfour}{$\mathrm{O} \left( 2n \right)/\mathrm{O} \left( n \right) \times \mathrm{O} \left( n \right)$}
\newcommand{\Rfive}{$\mathrm{U} \left( 2n \right)/\mathrm{O} \left( 2n \right)$}
\newcommand{\Rsix}{$\mathrm{Sp} \left( n \right)/\mathrm{U} \left( n \right)$}
\newcommand{\Rseven}{$\mathrm{Sp} \left( n \right)$}

% Page header
\markboth{Kawabata and Ryu}{Non-Hermitian Disordered Systems}

% Title
%\title{Field Theory of Disordered Non-Hermitian Systems} 
\title{Non-Hermitian Disordered Systems} 
%Title: Subtitle

%Authors, affiliations address.
\author{Kohei Kawabata$^1$ and Shinsei Ryu$^2$
\affil{$^1$Institute for Solid State Physics, University of Tokyo, Kashiwa, Chiba 277-8581, Japan; \\
email: kawabata@issp.u-tokyo.ac.jp}
\affil{$^2$Department of Physics, Princeton University, Princeton, New Jersey, 08544, USA; \\
email: shinseir@princeton.edu}}

%Abstract
\begin{abstract}
%Abstract text, approximately 150 words. 
Non-Hermitian disordered systems have emerged as a central arena in modern physics, with ramifications spanning condensed matter, quantum, statistical, and high energy contexts. 
The same principles also underlie phenomena beyond physics, such as network science, complex systems, and biophysics, where dissipation, nonreciprocity, and stochasticity are ubiquitous.
Here, we review the physics and mathematics of non-Hermitian disordered systems,
with particular emphasis on non-Hermitian random matrix theory.
We begin by presenting the 38-fold symmetry classification of non-Hermitian systems, 
contrasting it with the 10-fold way for Hermitian systems.
After introducing the classic Ginibre ensembles of non-Hermitian random matrices, we survey various diagnostics for complex-spectral statistics and distinct universality classes realized by symmetry.
As a key application to physics, we discuss how non-Hermitian random matrix theory characterizes chaos and integrability in open quantum systems.
We then turn to the criticality due to the interplay of disorder and non-Hermiticity, including Anderson transitions in the Hatano-Nelson model and its higher-dimensional extensions.
We also discuss the effective field theory description of non-Hermitian disordered systems in terms of nonlinear sigma models.
\end{abstract}

%Keywords, etc.
\maketitle

%Table of Contents
\tableofcontents

% Heading 1
\section{INTRODUCTION}
%Please begin the main text of your article here. 

%{\color{red}
%Total words (original text): 1042
%\\
%Word count excluding references/citations: 1003 words}

Non-Hermitian operators have become an indispensable language for modern condensed matter physics and related fields~\cite{Konotop-review, Christodoulides-review}. 
In practice, quantum dynamics is rarely perfectly closed: 
coupling to environments, engineered dissipation, and continuous monitoring all generate effectively nonunitary evolution~\cite{Nielsen-textbook, Breuer-textbook, Rivas-textbook}.  
Disorder is equally unavoidable, arising from impurities, fabrication imperfections, spatial inhomogeneities, and noisy control.  
The interplay of these two general ingredients---non-Hermiticity and disorder---yields a regime where familiar organizing principles must be reconsidered, and where new forms of universality emerge. 
In this review, we elucidate the physics and mathematics of non-Hermitian disordered systems, with particular emphasis on the fundamental roles of symmetry and universality.

A defining feature of non-Hermitian systems is that their spectra are generally complex, thereby reshaping the notion of spectral correlations.
When eigenvalues spread over the two-dimensional complex plane, many of the standard tools devised for real spectra cease to apply, and the choice of diagnostics for complex-spectral statistics becomes a matter of physics rather than mere technicality.
These issues are not purely formal:
complex spectra simultaneously encode decay rates and oscillation frequencies in open quantum dynamics and are directly relevant to experimentally accessible platforms, such as photonic structures and electric circuits with gain, loss, or nonreciprocity, as well as cold-atom and solid-state settings with controlled dissipation.
In high energy physics, non-Hermitian operators likewise arise naturally, for example, in Dirac operators at finite chemical potential~\cite{Markum-99, Akemann-02, Akemann-04, Osborn-04}.
More broadly, they provide effective descriptions across the natural sciences, from network science and population dynamics to biophysics~\cite{May-72, Nelson-98, Amir-16}.

The central viewpoint of this review is that symmetry and universality remain the proper guiding principles, but that these principles themselves must be upgraded beyond the Hermitian regime.
The celebrated 10-fold way generally classifies Hermitian systems in terms of discrete symmetry of time reversal, particle-hole transformation, and chiral transformation~\cite{Wigner-58, Dyson-62, AZ-97}. 
In non-Hermitian settings, however, it becomes essential to distinguish constraints acting on an operator from those acting on its Hermitian conjugate, 
so that symmetry notions that coincide in the Hermitian regime can bifurcate into genuinely inequivalent forms.
This enlargement culminates in the 38-fold symmetry classification~\cite{Bernard-LeClair-02, KSUS-19}, which provides a unified framework for organizing non-Hermitian systems and their universal phenomena. 

As in the Hermitian case, non-Hermitian random matrix theory presents canonical model settings for sufficiently complicated systems with complex spectra~\cite{Ginibre-65, Girko-85, Sommers-88, Grobe-89, Lehmann-91, Feinberg-97, Chalker-98, Nishigaki-02, Bernard-LeClair-02, Kanzieper-05, Forrester-07, KSUS-19, Hamazaki-20, Cipolloni-21, Akemann-22, Xiao-22, Kawabata-23SVD, Xiao-24HE, Akemann-25, Kulkarni-25, Forrester-25, Mehta-textbook, Byun-Forrester-textbook}. 
The classic Ginibre ensembles~\cite{Ginibre-65} already capture the basic phenomenon that generic non-Hermitian spectra fill two-dimensional domains in the complex plane, yet the modern perspective is broader: 
once symmetry is incorporated systematically, non-Hermitian universality extends far beyond Ginibre~\cite{Hamazaki-20}. 
At the same time, analytic control is still limited only for a handful of foundational symmetry classes, and even basic questions about universal correlations and crossover behavior are actively developing. 
A practical issue is how to diagnose universal correlations in complex spectra.
In Hermitian systems, nearest-neighbor level-spacing statistics provide a standard probe of local spectral correlations.
For complex spectra, however, a naive transplant of the one-dimensional notions fails. 
We therefore survey a variety of diagnostics for complex-spectral statistics that are both conceptually meaningful and practically useful.

One of the most timely motivations comes from open quantum many-body physics and the quest to extend the notion of quantum chaos beyond closed systems~\cite{Grobe-88, Haake-92, Hamazaki-19, Denisov-19, Can-19PRL, Can-19JPhysA, Sa-20JPhysA, Akemann-19, Sa-20, Wang-20, Xu-21, JiachenLi-21, Tarnowski-21, GarciaGarcia-22PRX, Sa-22SYK, Kulkarni-22SYK, GJ-23, Shivam-22, Ghosh-22, Costa-23, Sa-23, Kawabata-23}. 
In closed quantum systems, random matrix theory underlies the fundamental distinction between chaotic and integrable dynamics through universal spectral correlations~\cite{Berry-Tabor-77, BGS-84, Huse-review, Rigol-review, Haake-textbook}. 
In open quantum systems, by contrast, effective generators, such as non-Hermitian Hamiltonians and Lindblad superoperators, arise ubiquitously, 
yet the very definition of quantum chaos is less sharp because nonunitarity qualitatively changes the dynamics.
Complex-spectral statistics provide powerful diagnostics, 
but a broadly accepted formulation of chaos for open quantum systems remains an outstanding challenge. 
This stands in contrast to classical chaos, whose significance extends equally to open and closed settings.
For example, the Lorenz equation~\cite{Lorenz-63}, a paradigmatic example of chaotic dynamics, is a set of nonlinear differential equations for a dissipative system.
Still, placing the spectral diagnostics within the symmetry-guided framework of non-Hermitian universality, one can clarify which signatures are robust or model-dependent, and what kinds of theoretical structures are still missing.

Disorder further enriches the landscape by linking non-Hermitian physics to localization and criticality~\cite{Hatano-Nelson-96, Efetov-97, Brouwer-97, Hatano-Nelson-97, Efetov-97B, Feinberg-99, Goldsheid-98, Mudry-98, Hatano-Nelson-98, Longhi-15, Gong-18, Longhi-19, Zeng-20, Tzortzakakis-20, Huang-20, KR-21, Claes-21, Luo-21L, Luo-21B, Luo-22R, Liu-Fulga-21, Kim-21, Moustaj-22, DeTomasi-22, DeTomasi-23, Nakai-24, Ghosh-23, Thompson-24, Chen-25, Li-25}. 
In Hermitian systems, disorder induces metal-insulator transitions known as Anderson transitions~\cite{Anderson-58, Abrahams-79, Evers-review}. 
Non-Hermiticity changes this picture, 
sometimes by enabling qualitatively new delocalization mechanisms and sometimes by quantitatively changing universal data within given symmetry classes.
Paradigmatic examples, most notably the Hatano-Nelson model~\cite{Hatano-Nelson-96, Hatano-Nelson-97}, reveal delocalization mechanisms without direct Hermitian counterparts, such as intrinsic non-Hermitian topology~\cite{Gong-18, KSUS-19, Bergholtz-review, Okuma-Sato-review}, and critical behavior that can depart from the standard universality classes.

To unify the random matrix universality, dissipative quantum chaos, and disorder-driven criticality, an effective field theory perspective is significant. 
In Hermitian disordered systems, the nonlinear sigma model provides a systematic framework for diffusion, localization, and universal spectral correlations~\cite{Efetov-textbook, Evers-review, Haake-textbook, Kamenev-textbook}. 
In non-Hermitian settings, nonlinear sigma model descriptions remain equally powerful, but require conceptual refinements to handle genuinely two-dimensional complex spectra, where conventional constructions tied to real spectra must be reworked~\cite{Efetov-97, Efetov-97B, Nishigaki-02, Chen-25}. 
We review how nonlinear sigma models arise for non-Hermitian disordered systems and how they organize the universality classes, while also highlighting directions where field theory may help bridge current gaps.

This review is organized as follows.  
In Sec.~\ref{sec: symmetry}, we review the 10-fold way for Hermitian systems and then explain how non-Hermiticity enlarges it to the 38-fold way.  
In Sec.~\ref{sec: NH RMT}, we introduce non-Hermitian random matrix theory, beginning with the classic Ginibre ensembles and then discussing various diagnostics of complex-spectral statistics and distinct universality classes realized by symmetry. 
We also present nonlinear sigma models as an effective field theory framework. 
In Sec.~\ref{sec: dissipative quantum chaos}, we examine dissipative quantum chaos and clarify what complex-spectral statistics can, and cannot yet, establish about chaotic versus integrable regimes in open quantum systems.  
In Sec.~\ref{sec: Anderson}, we discuss disorder-induced criticality, Anderson transitions, unique to non-Hermitian disordered systems. 
In Sec.~\ref{sec: discussion}, we conclude with outlooks and open problems.

%%%%%%%%%%%%%%%%%%%%%%%%
\section{SYMMETRY}
    \label{sec: symmetry}

%{\color{red}Total word count: 1,024 words}

Symmetry underlies the distinct universality classes of disordered systems.
We review the 10-fold symmetry classification of Hermitian operators (Table~\ref{tab: Hermitian})~\cite{AZ-97} and the 38-fold symmetry classification of non-Hermitian operators (Tables~\ref{tab: complex AZ}, \ref{tab: real AZ}, \ref{tab: real AZ-dag}, and \ref{tab: real AZ + SLS})~\cite{KSUS-19}.

%%%%% Hermitian %%%%%
\begin{table}[t]
    \tabcolsep4pt
	\centering
	\caption{10-fold Altland-Zirnbauer (AZ) symmetry classes of Hermitian systems based on time-reversal symmetry (TRS), particle-hole symmetry (PHS), and chiral symmetry (CS).
    The fermionic replica nonlinear sigma model (NLSM) target manifolds, and possible topological terms in one, two, and three dimensions are also shown.}
	\label{tab: Hermitian}
     \begin{tabular}{cccccccc} \hline \hline
    ~~Class~~ & ~~TRS~~ & ~~PHS~~ & ~~CS~~ & ~~NLSM target manifold~~ & ~~$d=1$~~ & ~~$d=2$~~ & ~~$d=3$~~ \\ \hline
    A & $0$ & $0$ & $0$ & \Czero & $0$ & \Z & $0$ \\
    AIII & $0$ & $0$ & $1$ & \Cone & \Z & $0$ & \Z \\ \hline
    AI & $+1$ & $0$ & $0$ & \Rzero & $0$ & $0$ & $0$ \\
    BDI & $+1$ & $+1$ & $1$ & \Rone & \Z & $0$ & $0$ \\
    D & $0$ & $+1$ & $0$ & \Rtwo & \Ztwo & \Z & $0$ \\
    DIII & $-$ & $+$ & $1$ & \Rthree & \Ztwo & \Ztwo & \Z \\
    AII & $-1$ & $0$ & $0$ & \Rfour & $0$ & \Ztwo & \Ztwo \\
    CII & $-1$ & $-1$ & $1$ & \Rfive & \twoZ & $0$ & \Ztwo \\
    C & $0$ & $-1$ & $0$ & \Rsix & $0$ & \twoZ & $0$ \\
    CI & $+1$ & $-1$ & $1$ & \Rseven & $0$ & $0$ & \twoZ \\ \hline \hline
  \end{tabular}
\end{table}

%%%%% complex AZ %%%%%
\begin{table}[t]
    \tabcolsep0pt
    \centering
	\caption{Complex Altland-Zirnbauer (AZ) symmetry classes for non-Hermitian systems based on chiral symmetry (CS) and sublattice symmetry (SLS).
    The subscript of SLS $\mathcal{S}_{\pm}$ specifies the commutation ($+$) or anticommutation ($-$) relation to CS: $\Gamma \mathcal{S}_{\pm} = \pm \mathcal{S}_{\pm} \Gamma$.
    The Hermitized symmetry classes, fermionic replica nonlinear sigma model (NLSM) target manifolds, and possible topological terms in one, two, and three dimensions are also shown.
    }
	\label{tab: complex AZ}
     \begin{tabular}{cccccccc} \hline \hline
    ~~Class~~ & ~~CS~~ & ~~SLS~~ & ~~Hermitization~~ & ~~NLSM target manifold~~ & ~~$d=1$~~ & ~~$d=2$~~ & ~~$d=3$~~ \\ \hline
    A & $0$ & $0$ & AIII & \Cone & \Z & $0$ & \Z \\
    AIII = A + $\eta$ & $1$ & $0$ & A & \Czero & $0$ & \Z & $0$ \\ \hline
    AIII $+ \mathcal{S}_{+}$ & $1$ & $1$ & AIII & \Cone & \Z & $0$ & \Z \\ \hline
    ~~A $+ \mathcal{S}$ = AIII$^{\dag}$~~ & $0$ & $1$ & ~~AIII $\times$ AIII~~ & \Cone\,$\times$\,\Cone & \ZoplusZ & $0$ & \ZoplusZ \\
    AIII $+ \mathcal{S}_{-}$ & $1$ & $1$ & A $\times$ A & [\Czero]\,$\times$\,[\Czero] & $0$ & \ZoplusZ & $0$ \\ \hline \hline
  \end{tabular}
\end{table}

%%%%% real AZ %%%%%
\begin{table}[t]
	\centering
    \tabcolsep0pt
	\caption{Real Altland-Zirnbauer (AZ) symmetry classes for non-Hermitian systems based on time-reversal symmetry (TRS), particle-hole symmetry (PHS), and chiral symmetry (CS).
    The Hermitized symmetry classes, fermionic replica nonlinear sigma model (NLSM) target manifolds, and possible topological terms in one, two, and three dimensions are also shown.}
	\label{tab: real AZ}
     \begin{tabular}{ccccccccc} \hline \hline
    ~~Class~~ & ~~TRS~~ & ~~PHS~~ & ~~CS~~ & ~~Hermitization~~ & ~~NLSM target manifold~~ & ~~$d=1$~~ & ~~$d=2$~~ & ~~$d=3$~~ \\ \hline
    ~~AI = D$^{\dag}$~~ & +1 & 0 & 0 & BDI & \Rone & \Z & $0$ & $0$ \\
    BDI & +1 & +1 & 1 & D & \Rtwo & \Ztwo & \Z & $0$ \\
    D & 0 & +1 & 0 & DIII & \Rthree & \Ztwo & \Ztwo & \Z \\
    DIII & -1 & +1 & 1 & AII & \Rfour & $0$ & \Ztwo & \Ztwo \\
    AII = C$^{\dag}$ & -1 & 0 & 0 & CII & \Rfive & \twoZ & $0$ & \Ztwo \\
    CII & -1 & -1 & 1 & C & \Rsix & $0$ & \twoZ & $0$ \\
    C & 0 & -1 & 0 & CI & \Rseven & $0$ & $0$ & \twoZ \\
    CI & +1 & -1 & 1 & AI & \Rzero & $0$ & $0$ & $0$ \\ \hline \hline
  \end{tabular}
\end{table}

%%%%% real AZ-dag %%%%%
\begin{table}[t]
    \tabcolsep0pt
    \centering
	\caption{Real Altland-Zirnbauer$^{\dag}$ (AZ$^{\dag}$) symmetry classes for non-Hermitian systems based on time-reversal symmetry$^{\dag}$ (TRS$^{\dag}$), particle-hole symmetry$^{\dag}$ (PHS$^{\dag}$), and chiral symmetry (CS). 
    The Hermitized symmetry classes, fermionic replica nonlinear sigma model (NLSM) target manifolds, and possible topological terms in one, two, and three dimensions are also shown.}
	\label{tab: real AZ-dag}
     \begin{tabular}{ccccccccc} \hline \hline
    ~~Class~~ & ~~TRS$^{\dag}$~~ & ~~PHS$^{\dag}$~~ & ~~CS~~ & ~~Hermitization~~ & ~~NLSM target manifold~~ & ~~$d=1$~~ & ~~$d=2$~~ & ~~$d=3$~~ \\ \hline
    AI$^{\dag}$ & +1 & 0 & 0 & CI & \Rseven & $0$ & $0$ & \twoZ \\
    BDI$^{\dag}$ & +1 & +1 & 1 & AI & \Rzero & $0$ & $0$ & $0$ \\
    D$^{\dag}$ = AI & 0 & +1 & 0 & BDI & \Rone & \Z & $0$ & $0$ \\
    DIII$^{\dag}$ & -1 & +1 & 1 & D & \Rtwo & \Ztwo & \Z & $0$ \\
    AII$^{\dag}$ & -1 & 0 & 0 & DIII & \Rthree & \Ztwo & \Ztwo & $0$ \\
    CII$^{\dag}$ & -1 & -1 & 1 & AII & \Rfour & $0$ & \Ztwo & \Ztwo \\
    C$^{\dag}$ = AII & 0 & -1 & 0 & CII & \Rfive & \twoZ & $0$ & \Ztwo \\
    CI$^{\dag}$ & +1 & -1 & 1 & C & \Rsix & $0$ & \twoZ & $0$ \\ \hline \hline
  \end{tabular}
\end{table}

%%%%% real AZ + SLS %%%%%
\begin{table}[t]
    \tabcolsep0pt
	\centering
	\caption{Real Altland-Zirnbauer (AZ) symmetry classes with sublattice symmetry (SLS).
    The subscript of SLS $\mathcal{S}_{\pm}$ specifies the commutation ($+$) or anticommutation ($-$) relation to time-reversal symmetry (TRS) and/or particle-hole symmetry (PHS).
    For the symmetry classes that involve both TRS and PHS (i.e., classes BDI, DIII, CII, and CI), the first subscript specifies the relation to TRS and the second one to PHS.
    The Hermitized symmetry classes, fermionic replica nonlinear sigma model (NLSM) target manifolds, and possible topological terms in one, two, and three dimensions are also shown.
    Some classes are repeated to clarify the periodicity of the table.}
	\label{tab: real AZ + SLS}
     \begin{tabular}{cccccc} \hline \hline
    ~~Class~~ & ~~Hermitization~~ & ~~NLSM target manifold~~ & ~~$d=1$~~ & ~~$d=2$~~ & ~~$d=3$~~ \\ \hline
    BDI + $\mathcal{S}_{++}$ & BDI & \Rone & \Z & $0$ & $0$  \\
    DIII + $\mathcal{S}_{--}$ = BDI + $\mathcal{S}_{--}$ & DIII & \Rthree & \Ztwo & \Ztwo & \Z \\
    CII + $\mathcal{S}_{++}$ & CII & \Rfive & \twoZ & $0$ & \Ztwo \\
    CI + $\mathcal{S}_{--}$ = CII + $\mathcal{S}_{--}$ & CI & \Rseven & $0$ & $0$ & \twoZ \\ \hline
    AI + $\mathcal{S}_{-}$ = AII + $\mathcal{S}_{-}$ & AIII & \Cone & \Z & $0$ & \Z \\
    BDI + $\mathcal{S}_{-+}$ = DIII + $\mathcal{S}_{-+}$ & A & \Czero & $0$ & \Z & $0$ \\
    D + $\mathcal{S}_{+}$ & AIII & \Cone & \Z & $0$ & \Z \\
    DIII + $\mathcal{S}_{-+}$ = BDI + $\mathcal{S}_{-+}$ & A & \Czero & $0$ & \Z & $0$ \\
    AII + $\mathcal{S}_{-}$ = AI + $\mathcal{S}_{-}$ & AIII & \Cone & \Z & $0$ & \Z \\
    CII + $\mathcal{S}_{-+}$ = CI + $\mathcal{S}_{-+}$ & A & \Czero & $0$ & \Z & $0$ \\
    C + $\mathcal{S}_{+}$ & AIII & \Cone & \Z & $0$ & \Z \\
    CI + $\mathcal{S}_{-+}$ = CII + $\mathcal{S}_{-+}$ & A & \Czero & $0$ & \Z & $0$ \\ \hline
    BDI + $\mathcal{S}_{--}$ = DIII + $\mathcal{S}_{--}$ & DIII & \Rthree & \Ztwo & \Ztwo & \Z \\
    DIII + $\mathcal{S}_{++}$ & CII & \Rfive & \twoZ & $0$ & \Ztwo \\
    CII + $\mathcal{S}_{--}$ = CI + $\mathcal{S}_{--}$ & CI & \Rseven & $0$ & $0$ & \twoZ \\
    CI + $\mathcal{S}_{++}$ & BDI & \Rone & \Z & $0$ & $0$ \\ \hline
    AI + $\mathcal{S}_{+}$ & ~~BDI $\times$ BDI~~ & [\Rone]\,$\times$\,[\Rone] & \ZoplusZ & $0$ & $0$ \\
    BDI + $\mathcal{S}_{+-}$ & ~~D $\times$ D~~& [\Rtwo]\,$\times$\,[\Rtwo] & \ZtwooplusZtwo & \ZoplusZ & $0$ \\
    D + $\mathcal{S}_{-}$ & ~~DIII $\times$ DIII~~ & \Rthree\,$\times$\,\Rthree & \ZtwooplusZtwo & \ZtwooplusZtwo & \ZoplusZ \\
    DIII + $\mathcal{S}_{+-}$ & ~~AII $\times$ AII~~ & [\Rfour]\,$\times$\,[\Rfour] & $0$ & \ZtwooplusZtwo & \ZtwooplusZtwo \\
    AII + $\mathcal{S}_{+}$ & ~~CII $\times$ CII~~ & [\Rfive]\,$\times$\,[\Rfive] & \twoZoplustwoZ & $0$ & \ZtwooplusZtwo \\
    CII + $\mathcal{S}_{+-}$ & ~~C $\times$ C~~ & [\Rsix]\,$\times$\,[\Rsix] & $0$ & \twoZoplustwoZ & $0$ \\
    C + $\mathcal{S}_{-}$ & ~~CI $\times$ CI~~ & \Rseven\,$\times$\,\Rseven & $0$ & $0$ & \twoZoplustwoZ \\
    CI + $\mathcal{S}_{+-}$ & ~~AI $\times$ AI~~ & [\Rzero]\,$\times$\,[\Rzero] & $0$ & $0$ & $0$\\ \hline \hline
  \end{tabular}
\end{table}

%%%%%%%%%%%%
\subsection{10-fold way for Hermitian systems}

In general, Hermitian operators $H$ are classified according to the presence or absence of two types of antiunitary symmetry: time-reversal symmetry (TRS),
\begin{equation}
    \mathcal{T} H \mathcal{T}^{-1} = H, \quad \mathcal{T}^2 =\pm 1,
        \label{eq: TRS}
\end{equation}
and particle–hole symmetry (PHS),
\begin{equation}
    \mathcal{C} H \mathcal{C}^{-1} = - H, \quad \mathcal{C}^2 = \pm 1.
        \label{eq: PHS-dag}
\end{equation}
Here, $\mathcal{T}$ and $\mathcal{C}$ are antiunitary operators: 
$\mathcal{T}$ satisfies $\mathcal{T}^{\dag} \mathcal{T} = \mathcal{T} \mathcal{T}^{\dag} = 1$ and involves complex conjugation (i.e., $\mathcal{T} z \mathcal{T}^{-1} = z^{*}$ for $z \in \mathbb{C}$). 
As a combination of TRS and PHS, we further introduce chiral symmetry (CS), or equivalently sublattice symmetry (SLS), defined by
\begin{equation}
    \mathcal{S} H \mathcal{S}^{-1} = - H, \quad \mathcal{S}^2 = 1,
        \label{eq: SLS}
\end{equation}
where $\mathcal{S}$ is a unitary operator.
These two antiunitary symmetries together with one unitary symmetry constitute the 10-fold Altland-Zirnbauer (AZ) symmetry classification (Table~\ref{tab: Hermitian})~\cite{AZ-97}.

Within the 10 symmetry classes, TRS alone yields the Wigner-Dyson class (classes A, AI, and AII)~\cite{Wigner-58, Dyson-62}.
PHS underlies the physics of superconductors and superfluids, giving rise to the Bogoliubov-de Gennes class (classes D, DIII, CII, C, and CI)~\cite{AZ-97}.
CS characterizes the remaining symmetry classes (classes AIII, BDI, and CII) and is relevant to, for example, the sublattice structure~\cite{Gade-93} and quantum chromodynamics~\cite{Verbaarschot-94}.
Class AIII also appears in Bogoliubov-de Gennes systems with spin conservation and TRS.

The AZ classification exhausts all the possible discrete symmetries acting on the internal degrees of freedom, while excluding spatial symmetries, such as reflection, rotation, and inversion symmetries.
Consequently, as discussed further below, it governs the physics of disordered systems and determines the universality classes of Hermitian random matrices and Anderson transitions~\cite{Evers-review}, as well as topological insulators and superconductors~\cite{CTSR-review}.
In Table~\ref{tab: Hermitian}, we also list the topological classification of band insulators and Bogoliubov-de Gennes superconductors in one, two, and three spatial dimensions.
We assume that $H$ has no additional unitary symmetry commuting with it (i.e., $\mathcal{U} H \mathcal{U}^{-1} = H$); 
if such a symmetry is present, we block-diagonalize $H$ and then analyze the internal symmetry in each subspace.
Hence, such commutative unitary symmetry is not incorporated into the AZ classification.

%%%%% symmetry & spectra %%%%%
\begin{table}[t]
    \tabcolsep0pt
	\centering
	\caption{38-fold symmetry classification of non-Hermitian systems.
    Symmetry-imposed pair structures of complex eigenvalues and symmetry-preserving domains in the complex spectrum are shown.}
	\label{tab: symmetry & spectrum}
     \begin{tabular}{cccc} \hline \hline
    ~~Symmetry~~ & ~~Equation~~ & ~~Pairs~~ & ~~Symmetry-preserving domain~~ \\ \hline
    Time-reversal symmetry (TRS) & \eqref{eq: TRS} & $\left( E, E^{*} \right)$ & Real axis ($E \in \mathbb{R}$) \\
    ~~Time-reversal symmetry$^{\dag}$ (TRS$^{\dag}$)~~ & \eqref{eq: TRS-dag} & No & Anywhere ($E \in \mathbb{C}$) \\
    Particle-hole symmetry (PHS) & \eqref{eq: PHS} & $\left( E, -E \right)$ & Origin ($E = 0$) \\
    ~~Particle-hole symmetry$^{\dag}$ (PHS$^{\dag}$)~~ & \eqref{eq: PHS-dag} & ~~$\left( E, -E^{*} \right)$~~ & ~~Imaginary axis ($E \in \ii \mathbb{R}$)~~ \\
    Chiral symmetry (CS) & \eqref{eq: CS} & $\left( E, -E^{*} \right)$ & Imaginary axis ($E \in \ii \mathbb{R}$) \\
    Sublattice symmetry (SLS) & \eqref{eq: SLS} & $\left( E, -E \right)$ & Origin ($E = 0$) \\
    Pseudo-Hermiticity (pH) & \eqref{eq: pH} & $\left( E, E^{*} \right)$ & Real axis ($E \in \mathbb{R}$) \\ \hline \hline
    \end{tabular}
\end{table}

%%%%%%%%%%%%
\subsection{38-fold way for non-Hermitian systems}

In contrast to the 10-fold AZ symmetry classification for Hermitian operators, non-Hermitian operators are generally organized by the 38-fold internal symmetry scheme~\cite{Bernard-LeClair-02, KSUS-19}.
First, the two types of antiunitary symmetry in Eqs.~(\ref{eq: TRS}) and (\ref{eq: PHS-dag}) remain symmetry even for non-Hermitian operators $H$, each of which is denoted by TRS and PHS$^{\dag}$, respectively~\cite{KSUS-19}.
Likewise, the unitary symmetry in Eq.~(\ref{eq: SLS}) persists and is referred to as SLS.
In addition to these symmetries, non-Hermiticity allows for symmetry that relates non-Hermitian operators $H$ to their Hermitian conjugates $H^{\dag}$.
Specifically, we introduce two such antiunitary symmetries by
\begin{equation}
    \mathcal{T} H^{\dag} \mathcal{T}^{-1} = H, \quad \mathcal{T}^2 = \pm 1,
        \label{eq: TRS-dag}
\end{equation}
and
\begin{equation}
    \mathcal{C} H^{\dag} \mathcal{C}^{-1} = - H, \quad \mathcal{C}^2 = \pm 1,
        \label{eq: PHS}
\end{equation}
where $\mathcal{T}$ and $\mathcal{C}$ are antiunitary operators.
These symmetries are denoted by time-reversal symmetry$^{\dag}$ (TRS$^{\dag}$) and particle-hole symmetry (PHS), respectively, because they are obtained from TRS and PHS$^{\dag}$ in Eqs.~(\ref{eq: TRS}) and (\ref{eq: PHS-dag}) by appending Hermitian conjugation.
Notably, although TRS and TRS$^{\dag}$, as well as PHS and PHS$^{\dag}$, coincide with each other for Hermitian operators, they are distinct for non-Hermitian operators.

In the same spirit, we also consider Eq.~(\ref{eq: SLS}) with additional Hermitian conjugation by
\begin{equation}
    \Gamma H^{\dag} \Gamma^{-1} = - H, \quad \Gamma^2 = 1,
        \label{eq: CS}
\end{equation}
where $\Gamma$ is a unitary operator.
While unitary symmetry in Eq.~(\ref{eq: SLS}) is termed SLS for non-Hermitian operators, unitary symmetry in Eq.~(\ref{eq: CS}) is termed CS (SLS can also be considered as CS$^{\dag}$)~\cite{KSUS-19}.
The combination of TRS and PHS, as well as the combination of TRS$^{\dag}$ and PHS$^{\dag}$, yields CS;
on the other hand, the combination of TRS and PHS$^{\dag}$, as well as the combination of TRS$^{\dag}$ and PHS, yields SLS.

In this manner, non-Hermiticity splits and enriches each discrete symmetry, thereby proliferating the number of symmetry classes.
Analogous to the 10-fold AZ symmetry classification for Hermitian operators, TRS in Eq.~(\ref{eq: TRS}), PHS in Eq.~(\ref{eq: PHS}), and CS in Eq.~(\ref{eq: CS}) define a 10-fold symmetry classification for non-Hermitian operators (Tables~\ref{tab: complex AZ} and \ref{tab: real AZ}).
Moreover, TRS$^{\dag}$ in Eq.~(\ref{eq: TRS-dag}), PHS$^{\dag}$ in Eq.~(\ref{eq: PHS-dag}), and CS in Eq.~(\ref{eq: CS}) generate another 10-fold symmetry classification, which is called the AZ$^{\dag}$ symmetry class for non-Hermitian operators (Table~\ref{tab: real AZ-dag}).
In both the AZ and AZ$^{\dag}$ symmetry classes, SLS in Eq.~(\ref{eq: SLS}) is absent;
once SLS is incorporated as additional symmetry (Tables~\ref{tab: complex AZ} and \ref{tab: real AZ + SLS}), we obtain the full 38-fold symmetry classification for non-Hermitian operators~\cite{KSUS-19}.
Each symmetry imposes characteristic constraints on the complex spectrum, as summarized in Table~\ref{tab: symmetry & spectrum}.
For example, TRS generally enforces complex-conjugate pairs $\left( E, E^{*} \right)$ of eigenvalues or allows strictly real eigenvalues $E \in \mathbb{R}$.
While TRS$^{\dag}$ does not induce such pair structures of complex eigenvalues, TRS$^{\dag}$ with $\mathcal{T}^2 = -1$ instead leads to the degeneracy of generic complex eigenvalues---non-Hermitian generalization of Kramers' theorem.

Importantly, distinct symmetries encode distinct physical phenomena.
For example, TRS corresponds to time-reversal invariance even in open systems, and its spontaneous breaking leads to unique critical phenomena~\cite{Bender-98, Bender-02}.
By contrast, TRS$^{\dag}$ makes the scattering problem with an incoming wave from the left equivalent to that with an incoming wave from the right, thereby representing reciprocity in open systems.
Accordingly, the different 10-fold ways introduced above play different physical roles.
For instance, whereas individual quantum trajectories of monitored free fermions are categorized by the 10-fold AZ classification (Tables~\ref{tab: complex AZ} and \ref{tab: real AZ})~\cite{Xiao-24}, their average dynamics described by the Lindblad master equation falls into the 10-fold AZ$^{\dag}$ classification (Tables~\ref{tab: complex AZ} and \ref{tab: real AZ-dag})~\cite{Lieu-20}.
Furthermore, the effective non-Hermitian self-energy within the Green's function (or equivalently, Feshbach) formalism~\cite{Moiseyev-textbook} 
% Feshbach-58, Feshbach-62
is generally classified according to the AZ$^{\dag}$ scheme rather than the AZ scheme~\cite{Hamanaka-24}.
In the following, we elaborate on further implications of these distinct symmetries for the physics of non-Hermitian disordered systems.

Some of the symmetry classes listed in Tables~\ref{tab: complex AZ}, \ref{tab: real AZ}, \ref{tab: real AZ-dag}, and \ref{tab: real AZ + SLS} are mutually equivalent and therefore not 
to be
double counted in the 38-fold symmetry classification.
For example, if a non-Hermitian operator $H$ respects TRS in Eq.~(\ref{eq: TRS}), another non-Hermitian operator $\ii H$ respects PHS$^{\dag}$ in Eq.~(\ref{eq: PHS-dag}), both of which yield essentially the same universality class.
Consequently, classes AI and AII are respectively equivalent to classes D$^{\dag}$ and C$^{\dag}$ and share the identical universal characteristics (see Tables~\ref{tab: real AZ} and \ref{tab: real AZ-dag}).
For example, non-Hermitian random matrices in classes AI and AII exhibit essentially the same universal spectral statistics as those in class D$^{\dag}$ and C$^{\dag}$, up to a $90^{\circ}$ rotation in the complex plane. 
In particular, Anderson transitions in these classes share the same critical exponents.
As another example, suppose that we have a non-Hermitian operator $H$ that obeys pseudo-Hermiticity (pH)~\cite{Mostafazadeh-02-1, Mostafazadeh-02-2}, 
\begin{equation}
    \eta H^{\dag} \eta^{-1} = H, \quad \eta^2 = 1, 
        \label{eq: pH}
\end{equation}
placing it in class A + $\eta$.
Then, another non-Hermitian operator $\ii H$ satisfies CS in Eq.~(\ref{eq: CS}), $\eta \left( \ii H \right) \eta^{-1} = - \left( \ii H \right)$, and thus belongs to class AIII.
The complete correspondence among symmetry classes is provided in Table~XIII of Ref.~\cite{KSUS-19}.

%%%%%%%%%%%%%%%%%%%%%%%%
\section{NON-HERMITIAN RANDOM MATRIX THEORY}
    \label{sec: NH RMT}

Random matrix theory provides simple yet universal model settings for disordered systems.
Central quantities of interest include the spectral density and level-spacing distribution.
In stark contrast to real spectra of Hermitian systems, eigenvalues of non-Hermitian systems are generally distributed on the two-dimensional complex plane, thereby enriching the spectral statistics.
We review the classic Ginibre ensembles of non-Hermitian random matrices in Sec.~\ref{sec: Ginibre}, and then introduce various recently-proposed diagnostics for complex-spectral statistics in Sec.~\ref{sec: diagnostics}.
In Sec.~\ref{sec: NH RMT symmetry}, we elucidate how each symmetry in the 38-fold way manifests itself in complex-spectral statistics.
Finally, in Sec.~\ref{sec: NLSM}, we discuss the effective field theory description of non-Hermitian disordered systems in terms of nonlinear sigma models.

%%%%%%%%%%%%
\subsection{Ginibre ensemble}
    \label{sec: Ginibre}

\begin{figure}[t]
\centering
\includegraphics[width=1.0\linewidth]{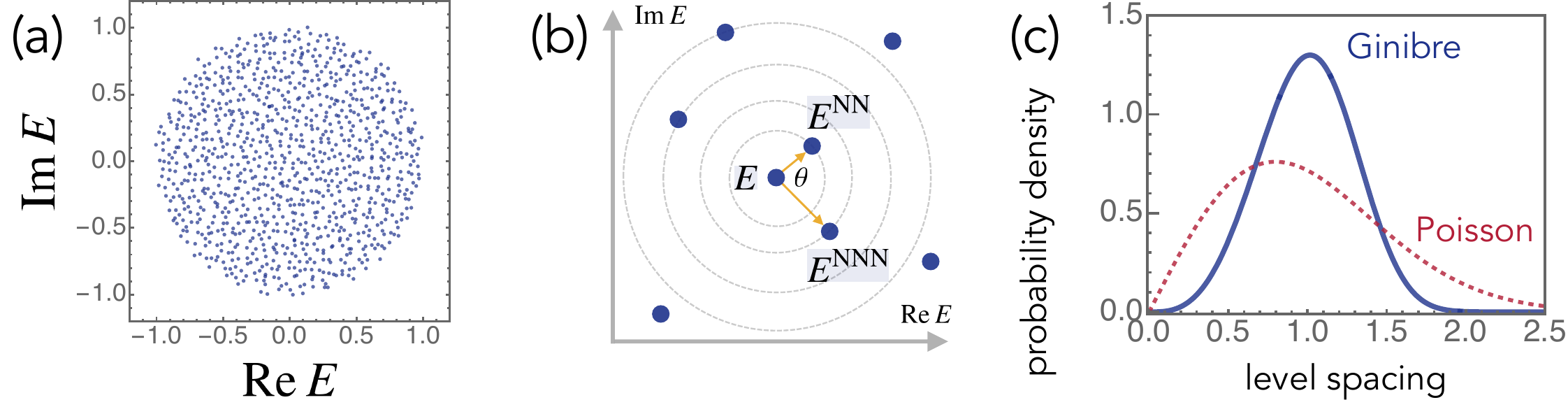}
\caption{Ginibre ensemble of non-Hermitian random matrices.
(a)~Complex spectrum of a 
$10^3 \times 10^3$
non-Hermitian random matrix drawn from the Ginibre unitary ensemble, with the spectral radius normalized to unity.
(b)~A complex eigenvalue $E$, its nearest-neighbor eigenvalue $E^{\text{NN}}$, and next-to-nearest-neighbor eigenvalue $E^{\text{NNN}}$.
(c)~Probability density of level spacings for the Ginibre ensemble (blue solid curve; Eq.~\eqref{eq: Ginibre level-spacing distribution}) and Poisson ensemble (red dashed curve; Eq.~\eqref{eq: Poisson}).
The level spacing is rescaled such that its mean is unity.}
    \label{fig: Ginibre}
\end{figure}

Classic ensembles of non-Hermitian random matrices were first introduced by Ginibre~\cite{Ginibre-65}.
The Ginibre unitary ensemble comprises $N \times N$ non-Hermitian random matrices whose entries are independent and identically distributed complex standard Gaussian variables.
In other words, matrices $H$ are drawn from the Gaussian measure proportional to $e^{- \mathrm{tr}\,H^{\dag}H}$.
This construction provides a non-Hermitian extension of the Gaussian unitary ensemble developed by Wigner~\cite{Wigner-58} 
% Wigner-51
and Dyson~\cite{Dyson-62}.
In Fig.~\ref{fig: Ginibre}\,(a), we show the complex spectrum of a single realization of such a non-Hermitian random matrix.
A remarkable feature is that the eigenvalues are supported on a disk with uniform density in the complex plane.
Indeed, in the large-$N$ limit, the spectral density $\rho \left( E \right)$ at $E \in \mathbb{C}$ obeys
\begin{equation}
    \rho \left( E \right) \coloneqq \frac{1}{N} \Braket{\sum_{n=1}^{N} \delta^{(2)} \left( E-E_n \right)} \sim \begin{cases}
        1/\pi & ( \left| E \right| < \sqrt{N} ); \\
        0 & ( \left| E \right| > \sqrt{N} ), \\
    \end{cases}
        \label{eq: Ginibre - DoS}
\end{equation}
where $E_n \in \mathbb{C}$ ($n=1, 2, \cdots, N$) is a complex eigenvalue for a single realization, and the bracket $\braket{\cdots}$ denotes the ensemble average.
This circular law (Girko's circular law~\cite{Girko-85}) is a non-Hermitian counterpart of Wigner's semicircle law in Hermitian random matrix theory.

The spectral density in Eq.~\eqref{eq: Ginibre - DoS} does not encode correlations among generic complex eigenvalues.
To probe spectral correlations, the distribution of level spacings provides a powerful diagnostic.
For each complex eigenvalue $E_n \in \mathbb{C}$, we define the level spacing by 
\begin{equation}
    s_n \coloneqq | E_n - E_{n}^{\rm NN} | \geq 0
        \label{eq: level spacing}
\end{equation}
with the nearest-neighbor eigenvalue $E_{n}^{\rm NN} \in \mathbb{C}$ in the complex plane [Fig.~\ref{fig: Ginibre}\,(b)].
This definition employs the full two-dimensional geometry of the spectrum.
By contrast, defining spacings solely from the real parts would generally fail to capture spectral correlations, since eigenvalues far apart in the complex plane may nevertheless have nearly identical real parts.

For the Ginibre unitary ensemble, the probability distribution function $P_{\rm Gin} \left( s \right)$ of the level spacing $s$ is exactly obtained as
\begin{equation}
    P_{\rm Gin} \left( s \right) = \left[ \prod_{k=1}^{\infty} \frac{\Gamma \left( 1+k, s^2 \right)}{k!}\right] \left[ \sum_{j=1}^{\infty} \frac{2s^{2j+1} e^{-s^2}}{\Gamma \left( 1+j, s^2 \right)} \right],
        \label{eq: Ginibre level-spacing distribution}
\end{equation}
with the incomplete Gamma function $\Gamma \left( 1+k, s^2 \right) \coloneqq \int_{s^2}^{\infty} t^k e^{-t} dt$ [Fig.~\ref{fig: Ginibre}\,(c)].
The mean level spacing
is $\int_{0}^{\infty} s P_{\rm Gin} \left( s \right) ds = 1.142929 \cdots$ [note that the spacing is rescaled to unit mean in Fig.~\ref{fig: Ginibre}\,(c)].
This probability distribution function grows as $P_{\rm Gin} \left( s \right) \propto s^3$ for $0 < s \ll 1$ and decays rapidly as $P_{\rm Gin} \left( s \right) \propto e^{-s^4/4}$ for $s \to \infty$.
For comparison, the Poisson distribution function of complex numbers, or equivalently, 
the probability distribution function of uncorrelated points in two dimensions, is given as
\begin{equation}
    P_{\rm P} \left( s \right) = \frac{\pi s}{2} e^{-\pi s^2/4},
        \label{eq: Poisson}
\end{equation}
where the level spacing is normalized such that its mean is unity\footnote{In passing, Eq.~\eqref{eq: Poisson} accidentally coincides with the level-spacing distribution of $2\times 2$ Hermitian random matrices in the Gaussian orthogonal ensemble (i.e., Wigner's surmise).}.
It should be noted that $P_{\rm P} \left( s \right)$ vanishes linearly for small level spacings, $P_{\rm P} \left( s \right) \propto s$ for $0 < s \ll 1$, which contrasts with its counterpart of real numbers, $P_{\rm P}^{\rm (real)} \left( s \right) = e^{-s} \to 1$ ($s \to 0$).
This linear decay arises merely from the two-dimensional Jacobian and does not signify the level repulsion.
Consequently, the cubic decay of the level-spacing distribution function $P_{\rm Gin} \left( s \right)$ provides a clear hallmark of the level repulsion between complex eigenvalues in non-Hermitian random matrices. 

In a spirit akin to Wigner-Dyson's 3-fold way~\cite{Dyson-62}, Ginibre also introduced ensembles of non-Hermitian random matrices with TRS in Eq.~(\ref{eq: TRS}).
Analogously to Gaussian unitary, orthogonal, and symplectic ensembles in Wigner-Dyson's 3-fold way, Ginibre's 3-fold ensembles of non-Hermitian random matrices are invariant under unitary, orthogonal, and symplectic transformations.
Specifically, Ginibre's 3-fold way is summarized as follows:
\begin{itemize}
    \item The Ginibre unitary ensemble (GinUE) consists of complex elements, as discussed above.
    These matrices possess no symmetry and therefore belong to class A in the 38-fold symmetry classification of Sec.~\ref{sec: symmetry}.

    \item The Ginibre orthogonal ensemble (GinOE) consists of real elements. 
    These matrices $H$ satisfy $H^{*} = H$ and hence respect TRS in Eq.~(\ref{eq: TRS}) with $\mathcal{T} = \mathcal{K}$, placing them in class AI.

    \item The Ginibre symplectic ensemble (GinSE) consists of quaternionic elements.
    Upon representing quaternions in terms of Pauli matrices $\sigma_{x, y, z}$, these matrices $H$ respect TRS in Eq.~(\ref{eq: TRS}) with $\mathcal{T} = \ii \sigma_y \mathcal{K}$ and thus belong to class AII.
\end{itemize}
As in the GinUE, generic complex eigenvalues obey the circular law in Eq.~\eqref{eq: Ginibre - DoS}
However, deviations arise around the real axis, on which TRS is preserved (see Table~\ref{tab: symmetry & spectrum}).
For the GinOE, the spectral density vanishes linearly as the real axis is approached, $\rho \left( E \right) \propto \left| \mathrm{Im}\,E\right|$, while a subextensive number of real eigenvalues appear.
For the GinSE, the spectral density vanishes quadratically toward the real axis, $\rho \left( E \right) \propto \left| \mathrm{Im}\,E\right|^2$, while no real eigenvalues appear in general.
On the other hand, the level-spacing distributions for the GinUE, GinOE, and GinSE coincide and are all given by Eq.~\eqref{eq: Ginibre level-spacing distribution}~\cite{Grobe-89}.
This stands in sharp contrast to Wigner-Dyson's 3-fold way, where TRS gives rise to three distinct universality classes of level-spacing statistics.
This apparent ``paradox" was recently resolved~\cite{Hamazaki-20} within the 38-fold symmetry classification of Sec.~\ref{sec: symmetry}; 
see Sec.~\ref{sec: NH RMT symmetry} for details.

The Ginibre ensembles have been widely employed to model sufficiently complicated open quantum systems and characterize nonintegrability and chaos in such settings (see Sec.~\ref{sec: dissipative quantum chaos} for details).
They have also been applied to capture the universal features of nonunitary dynamics subject to quantum measurements~\cite{Schomerus-22, Bulchandani-24, DeLuca-25, Xiao-25DMPK}.
In addition to such an open quantum regime, these random-matrix ensembles serve as effective models for intricate classical stochastic processes.
For example, they were invoked to analyze complex ecological systems~\cite{May-72}, where the circular law in Eq.~\eqref{eq: Ginibre - DoS} plays a central role in evaluating their stability.
Further mathematical details, as well as derivations of the aforementioned results, can be found in textbooks~\cite{Mehta-textbook, Haake-textbook, Byun-Forrester-textbook}.

%%%%%%%%%%%%
\subsection{Diagnostics of complex-spectral statistics}
    \label{sec: diagnostics}

\begin{figure}[t]
\centering
\includegraphics[width=1.0\linewidth]{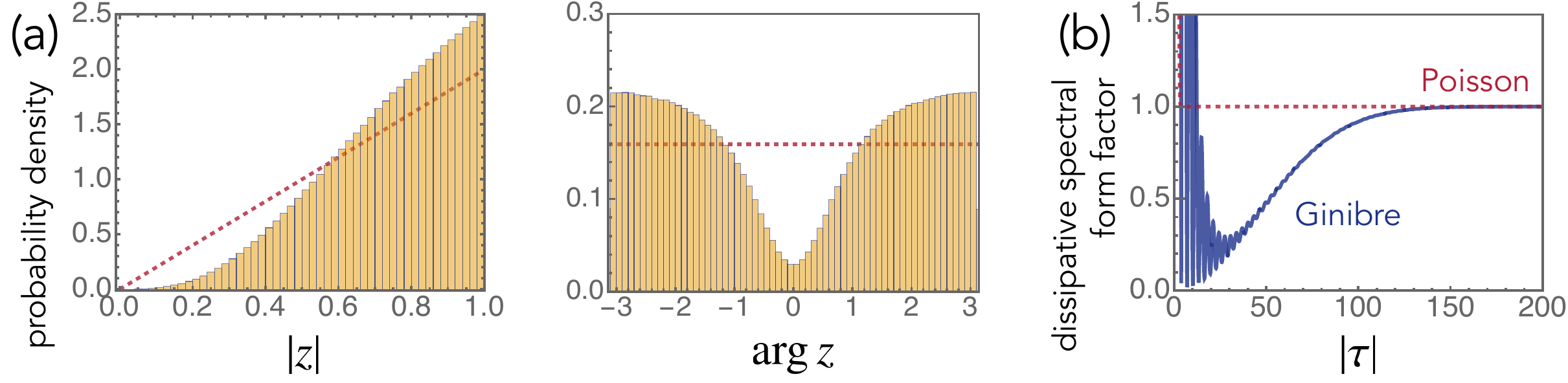}
\caption{Diagnostics of complex-spectral statistics.
(a)~Probability density of complex level-spacing ratios $z$ for their modulus $\left| z \right|$ and argument $\arg z$.
The numerical data are averaged over $10^4$ realizations of $10^3 \times 10^3$ matrices in the Ginibre unitary ensemble (Gaussian ensemble in class A).
Red dashed curve: Poisson ensemble [$P_{\rm P}^{\rm abs} \left( \left| z \right| \right) = 2 \left| z \right|$, $P_{\rm P}^{\rm arg} \left(  \arg z\right) = 1/2\pi$].
(b)~Dissipative spectral form factor $K \left( \tau, \tau^{*} \right)$ in Eq.~\eqref{eq: DSFF} as a function of the modulus $\left| \tau \right|$ of complex time $\tau$ ($N=10^3$).
Blue solid curve: Ginibre ensemble (class A) [$K_{\rm Gin} \left( \tau, \tau^{*} \right) \sim 1 + N \left( 2 J_1 \left( \left| \tau\right| \right)/\left| \tau \right| \right)^2 - e^{-\left| \tau\right|^2/4N}$ for $N \to \infty$].
Red dashed curve: Poisson ensemble [$K_{\rm P} \left( \tau, \tau^{*} \right) = 1 + \left( N-1 \right) e^{- \left| \tau \right|^2}$].}
    \label{fig: complex_spectral_statistics}
\end{figure}

Beyond the conventional diagnostics discussed in Sec.~\ref{sec: Ginibre}, several recent works have proposed alternative measures to quantify the spectral statistics of non-Hermitian disordered systems.
First, to characterize the level repulsion among complex eigenvalues, the statistics of level-spacing ratios have been introduced~\cite{Sa-20}:
\begin{equation}
    z_n \coloneqq \frac{E_{n}^{\rm NN} - E_n}{E_{n}^{\rm NNN} - E_n} \in \mathbb{C},
        \label{eq: level-spacing ratio}
\end{equation}
where $E_n^{\rm NN}$ and $E_{n}^{\rm NNN}$ denote the nearest-neighbor and next-to-nearest-neighbor eigenvalues for a complex eigenvalue $E_n$ [Fig.~\ref{fig: Ginibre}\,(b)].
The level spacing in Eq.~\eqref{eq: level spacing} has the dimension of energy and therefore depends on the local spectral density.
As a result, to make the level-spacing distribution a meaningful and comparable diagnostic, we need to unfold the complex spectrum to achieve a locally uniform density.
By contrast, the level-spacing ratio in Eq.~\eqref{eq: level-spacing ratio} is dimensionless and thus circumvents this additional technical procedure, analogous to the Hermitian counterpart~\cite{Oganesyan-07, Atas-13}.
Moreover, owing to the complex-valued nature of $z_n$, the statistics of both $\left| z_n \right|$ and $\arg z_n$ provide effective diagnostics of level repulsion [Fig.~\ref{fig: complex_spectral_statistics}\,(a)].
Specifically, level repulsion is reflected in the cubic suppression of $P^{\rm abs} \left( \left| z \right| \right)$ for $\left| z \right| \to 0$ and a pronounced dip of $P^{\rm arg} \left( \arg z \right)$ around $\arg z = 0$.

Another effective probe of spectral statistics in disordered systems is the spectral form factor.
For Hermitian systems, the spectral form factor is defined through the Fourier transform of the two-level correlation function $\braket{\rho \left( E \right) \rho \left( E+\omega \right)}$:
$K \left( t \right) \coloneqq 
N^{-1} \braket{| \sum_{n=1}^{N} e^{\ii E_n t} |^2}$~\cite{Haake-textbook}.
% Brezin-Hikami-97
In contrast to the level-spacing distribution function, the spectral form factor captures spectral correlations across all scales, encompassing both level repulsion and spectral rigidity.
A possible non-Hermitian extension of the spectral form factor, 
called the dissipative spectral form factor,
has been proposed as~\cite{JiachenLi-21}
\begin{equation}
    K \left( \tau, \tau^{*} \right) \coloneqq \frac{1}{N} \Braket{\left| \sum_{n=1}^{N} e^{\ii\,\mathrm{Re}\,(E_n \tau^{*})} \right|^2} = \frac{1}{N} \Braket{\sum_{m, n = 1}^{N} e^{\ii (\mathrm{Re}\,(E_m-E_n),\,\mathrm{Im}\,(E_m-E_n) ) \cdot (\mathrm{Re}\,\tau,\,\mathrm{Im}\,\tau)}},
        \label{eq: DSFF}
\end{equation}
where time $\tau \in \mathbb{C}$ is complexified as a consequence of complex energy $E \in \mathbb{C}$.
This is the two-dimensional Fourier transform of the two-level correlation function.
Like the Hermitian counterpart, the dissipative spectral form factor of the Ginibre ensemble exhibits a characteristic dip-ramp-plateau structure [Fig.~\ref{fig: complex_spectral_statistics}\,(b)]:
it initially decays, then grows quadratically up to the Heisenberg time $\tau_{\rm H} = \mathcal{O}\,( \sqrt{N} )$, and finally saturates at a plateau.
While the early-time decay can reflect nonuniversal features of the spectral density, the quadratic ramp originates from the spectral rigidity.
It should be noted that an alternative non-Hermitian generalization of the spectral form factor has also been proposed, which more faithfully captures the nonunitary dynamics but fails to encode the complex-spectral statistics~\cite{Can-19JPhysA, KKLNR-23}.

Whereas we have focused on the eigenvalue statistics, the statistics of eigenvectors also encode distinctive features of non-Hermitian disordered systems.
As a direct consequence of non-Hermiticity, right eigenstates $\ket{R_n}$'s ($H \ket{R_n} = E_n \ket{R_n}$) generally differ from left eigenstates $\ket{L_n}$'s ($H^{\dag} \ket{L_n} = E_n^{*} \ket{L_n}$)~\cite{Brody-14}.
In particular, eigenvector overlaps exhibit the statistics unique to non-Hermitian disordered systems~\cite{Chalker-98}:
\begin{equation}
    O_{mn} \coloneqq \braket{L_{m} | L_n} \braket{R_n | R_m},
\end{equation}
where we normalize the eigenvectors by $\braket{L_m | R_n} = \delta_{mn}$.
This quantity becomes trivial in Hermitian systems, $O_{mn} = \delta_{mn}$, yet quantifies nonorthogonality of eigenvectors in non-Hermitian systems.
The diagonal elements $O_{nn}$ control the spectral sensitivity of non-Hermitian matrices, known as the condition numbers in numerical analysis~\cite{Trefethen-Embree-textbook} and also as the Petermann factor~\cite{Petermann-79}.
Based on the statistics of eigenvector overlaps, recent works have investigated typical quantum entanglement~\cite{GJ-23} 
in non-Hermitian disordered systems.
Furthermore, the statistics of singular values likewise provide distinctive spectral diagnostics~\cite{Kawabata-23SVD}.

%%%%%%%%%%%%
\subsection{Distinct universality classes with symmetry}
    \label{sec: NH RMT symmetry}

\begin{figure}[t]
\centering
\includegraphics[width=1.0\linewidth]{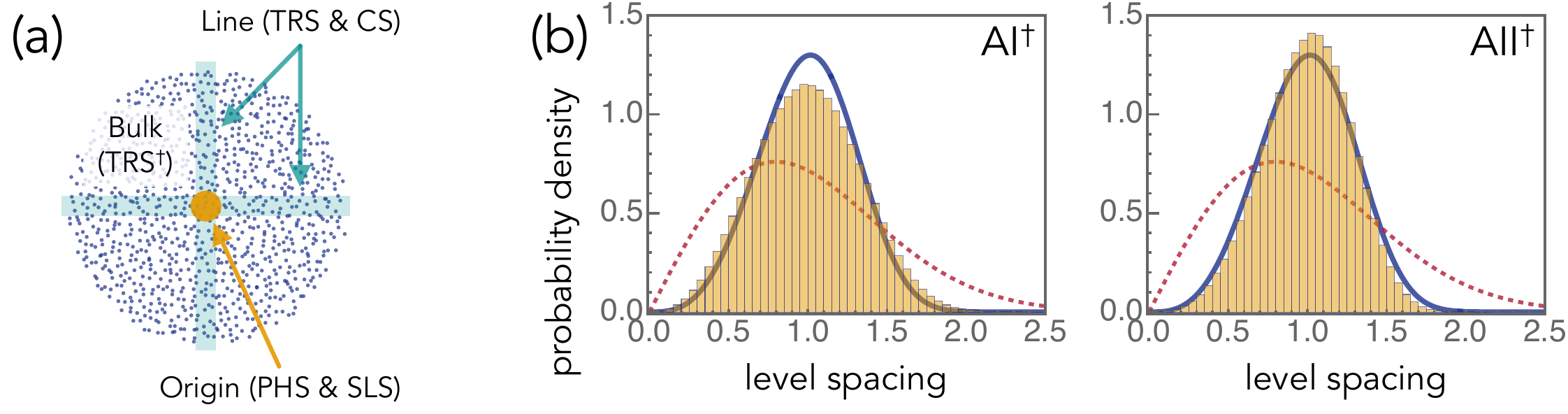}
\caption{Symmetry and non-Hermitian random matrix theory.
(a)~Spectral bulk, line (i.e., real or imaginary axis), and origin in the complex plane, where different symmetries are relevant.
(b)~Probability density of level spacings for non-Hermitian random matrices in class AI$^{\dag}$ and AII$^{\dag}$.
The numerical data are averaged over $10^4$ realizations of $10^3 \times 10^3$ matrices in the Gaussian ensemble.
Blue solid curve: Ginibre ensemble (class A; Eq.~\eqref{eq: Ginibre level-spacing distribution}).
Red dashed curve: Poisson ensemble (Eq.~\eqref{eq: Poisson}).}
    \label{fig: RMT_symmetry}
\end{figure}

In Hermitian random matrix theory, TRS changes spectral correlations of generic eigenvalues and gives rise to the three distinct universal level statistics, whereas PHS and CS affect correlations in the vicinity of the spectral origin.
More recently, the universality classes of the complex-spectral statistics have also been elucidated in non-Hermitian random matrix theory.
Each symmetry within the 38-fold classification of Sec.~\ref{sec: symmetry} is relevant to the spectral bulk, real or imaginary axis, and origin [Fig.~\ref{fig: RMT_symmetry}\,(a)], as discussed below.

%%%%%%
\subsubsection{Bulk}

As discussed in Sec.~\ref{sec: Ginibre}, the Ginibre ensembles based on TRS exhibit the common cubic level repulsion, $P \left( s \right) \propto s^3$ ($0 < s \ll 1$)~\cite{Grobe-89}.
This stands in sharp contrast to the Hermitian setting, where TRS changes the spectral correlations and yields the three distinct universality classes with $P \left( s \right) \propto s^{\beta}$ ($0 < s \ll 1$; $\beta = 1, 2, 4$)~\cite{Wigner-58, Dyson-62, Mehta-textbook, Haake-textbook}.
% Wigner-51
Notably, since complex conjugation and transposition are inequivalent operations in non-Hermitian systems, TRS in Eq.~\eqref{eq: TRS} and TRS$^{\dag}$ in Eq.~\eqref{eq: TRS-dag} are fundamentally distinct~\cite{KSUS-19}.
Moreover, whereas TRS is relevant only around the real axis, TRS$^{\dag}$ is preserved even for generic complex eigenvalues (see Table~\ref{tab: symmetry & spectrum}).

Motivated by these observations, it has recently been established that TRS$^{\dag}$, rather than TRS, changes the complex-spectral statistics and gives rise to the universal 3-fold level-spacing distributions [Fig.~\ref{fig: RMT_symmetry}\,(b)]~\cite{Hamazaki-20}, in close analogy with Wigner-Dyson's 3-fold way.
Even within this non-Hermitian 3-fold way, the level repulsion remains universally cubic, while class AI$^{\dag}$ accompanies a logarithmic correction, $P \left( s \right) \propto -s^3 \log s$ ($0 < s \ll 1$).
The influence of TRS$^{\dag}$ is instead encoded in the global shape of the level-spacing distribution $P \left( s \right)$:
the peak of $P \left( s \right)$ is suppressed in class AI$^{\dag}$ and enhanced in class AII$^{\dag}$, indicating weaker and stronger level repulsions for TRS$^{\dag}$ with $\mathcal{T}^2 = +1$ and $\mathcal{T}^2 = -1$, respectively.
In a similar spirit to the Wigner surmise, these generic features of TRS$^{\dag}$ are captured even by $2 \times 2$ or $4 \times 4$ non-Hermitian random matrices, for which $P \left( s \right)$ is exactly obtained as~\cite{Hamazaki-20}
\begin{equation}
    P \left( s \right) = \frac{\left( C_f s \right)^3}{\mathcal{N}_f} K_{\frac{f-2}{2}} \left( \left( C_f s \right)^2 \right) = \begin{cases}
        2C_2^4 s^3 K_0 \left( C_2^2 s^2 \right) & \left( \text{class~AI}^{\dag}; f=2\right); \\
        2C_3^4 s^3 e^{-C_3^2 s^2} & \left( \text{class~A}; f=3\right); \\
        \cfrac{2C_5^4 s^3}{3} \left( 1+C_5^2 s^2 \right) e^{-C_5^2 s^2} & \left( \text{class~AII}^{\dag}; f=5\right),
    \end{cases}
\end{equation}
with the modified Bessel function $K_{\alpha} \left( x \right)$, and the constants $C_2 = \Gamma^2 \left( 1/4 \right)/8\sqrt{2} = 1.16187 \cdots$, $C_3 = 3\sqrt{\pi}/4 = 1.32934 \cdots$, and $C_5 = 7\sqrt{\pi}/8 = 1.5509 \cdots$.
Despite substantial recent progress~\cite{Akemann-22, Akemann-25, Kulkarni-25, Forrester-25}, the exact level-spacing distributions for arbitrary $N$ in classes AI$^{\dag}$ and AII$^{\dag}$ remain unknown, constituting an outstanding mathematical challenge.
In passing, TRS$^{\dag}$ also plays a central role in Anderson transitions~\cite{KR-21} and skin effects~\cite{OKSS-20} of non-Hermitian systems, as discussed in Sec.~\ref{sec: Anderson}.

%%%%%%
\subsubsection{Real or imaginary axis}

As exemplified by Ginibre's 3-fold way, TRS changes the correlations in the vicinity of the real axis, resulting in the suppression of the spectral density and the emergence of a subextensive number of strictly real eigenvalues.
Similarly, pH in Eq.~\eqref{eq: pH} (PHS$^{\dag}$ in Eq.~\eqref{eq: PHS-dag} and CS in Eq.~\eqref{eq: CS}) is preserved only near the real (imaginary) axis.
In this manner, the universal spectral correlations around the real or imaginary axis are systematically classified, giving rise to the 10-fold AZ$^{\dag}$ symmetry classes in Tables~\ref{tab: complex AZ} and \ref{tab: real AZ-dag}~\cite{Xiao-22}.
These distinct universality classes are characterized by how the spectral density $\rho \left( E \right)$ vanishes toward the real axis:
\begin{equation}
    \rho \left( E \right) \propto \begin{cases}
        - \left| \mathrm{Im}\,E \right| \log \left| \mathrm{Im}\,E \right| & \left( \text{class~AI} + \eta_{+} \right); \\
        \left| \mathrm{Im}\,E \right| & \left( \text{class A} + \eta, \text{class~AI}, \text{class~AI} + \eta_{-}, \text{class~AII} + \eta_{\pm}\right); \\
        \left| \mathrm{Im}\,E \right|^2 & \left( \text{class~AII}\right). \\
    \end{cases}
        \label{eq: DoS - real axis}
\end{equation}
Additionally, the five symmetry classes (namely, class A + $\eta$, class AI, class AI + $\eta_{+}$, and classes AII + $\eta_{\pm}$) exhibit a subextensive number of real eigenvalues, whose spacings obey symmetry-specific universal distributions.

%%%%%%
\subsubsection{Origin}

The remaining symmetries, PHS in Eq.~\eqref{eq: PHS} and SLS in Eq.~\eqref{eq: SLS}, change the spectral correlations only in the vicinity of the origin $E=0$~\cite{Splittorff-04, Akemann-09, GarciaGarcia-22PRX, Xiao-24HE}.
The resulting universality classes are classified in terms of the decay of the spectral density $\rho \left( E \right)$ toward the origin.
Specifically, $\rho \left( E \right)$ follows
\begin{equation}
    \rho \left( E \right) \propto \begin{cases}
        \left| E \right| & \left( \text{class~D}, \text{class~D} + \mathcal{S}_{+}\right); \\
        - \left| E \right|^3 \log \left| E \right| & \left( \text{class~A} + \mathcal{S}, \text{class~C} + \mathcal{S}_{-}, \text{class~D} + \mathcal{S}_{-} \right); \\
        \left| E \right|^3 & \left( \text{class~C}, \text{class~C} + \mathcal{S}_{+}\right), \\
    \end{cases}
\end{equation}
for $E \to 0$.
Taken together, the complex-spectral statistics in the bulk, around the real or imaginary axis, and near the origin uniquely resolve the 38-fold universality classes of non-Hermitian random matrices.

%%%%%%%%%%%%
\subsection{Nonlinear sigma models}
    \label{sec: NLSM}

The effective field theory description of disordered systems is developed for both Hermitian~\cite{Efetov-textbook, Evers-review, Haake-textbook, Kamenev-textbook} and non-Hermitian~\cite{Efetov-97, Efetov-97B, Nishigaki-02, Chen-25} settings in terms of nonlinear sigma models.
In $d$ dimensions, the action is formulated by a matrix field $Q$ constrained to a target manifold and constructed from the simplest symmetry-allowed gradient terms,
\begin{equation}
    S [Q] = \frac{1}{2g} \int d^dx~\mathrm{tr} \left[ ( \nabla Q^{\dag} ) ( \nabla Q ) \right],
        \label{eq: NLSM}
\end{equation}
with the coupling constant $g > 0$.
In zero dimension, this nonlinear sigma model provides an effective description of random-matrix ensembles.
The target manifold of $Q$ depends on symmetry of the underlying microscopic disordered systems, as summarized in Table~\ref{tab: Hermitian} for Hermitian systems and Tables~\ref{tab: complex AZ}, \ref{tab: real AZ}, \ref{tab: real AZ-dag}, and \ref{tab: real AZ + SLS} for non-Hermitian systems.
For example, non-Hermitian systems in classes A, AI$^{\dag}$, and AII$^{\dag}$ correspond to target manifolds given by the unitary group $\mathrm{U} \left( n \right)$, symplectic group $\mathrm{Sp} \left( n \right)$, and orthogonal group $\mathrm{O} \left( n \right)$, respectively~\cite{Kulkarni-25}.

To elucidate the distinction between the Hermitian and non-Hermitian regimes, we review derivations of nonlinear sigma models from microscopic disordered systems within the replica formalism.
To capture the statistical properties of disordered systems $H$, it is useful to introduce the Green's function (equivalently, resolvent),
\begin{equation}
    G \left( E \right) \coloneqq \frac{1}{N} \braket{\mathrm{tr} \left( E-H \right)^{-1}}.
        \label{eq: Green}
\end{equation}
For Hermitian systems with real spectra $E \in \mathbb{R}$, the spectral density in Eq.~\eqref{eq: Ginibre - DoS} is obtained as $\rho \left( E \right) = - \pi^{-1} \lim_{\varepsilon \to 0^+} \mathrm{Im}\,G \left( E+\ii \varepsilon \right)$, where we use the formula of the one-dimensional delta function, $\lim_{\varepsilon \to 0^+} \mathrm{Im} \left( E+\ii \varepsilon \right)^{-1} = - \pi \delta \left( E \right)$.
Then, $G \left( E \right)$ is expressed as
\begin{equation}
    G \left( E \right) = \frac{1}{N} \lim_{n\to 0} \frac{1}{n} \frac{\partial}{\partial E} Z_n \left( E \right), \quad Z_n \left( E \right) \coloneqq \braket{\left[ \det \left( E-H \right)\right]^n}.
        \label{eq: replica-Hermitian}
\end{equation}
An advantage of the replica method is that the replica partition function $Z_n \left( E \right)$ ($n > 0$) admits a representation in terms of fermionic path integrals.
While averaging over Gaussian disorder generates quartic fermion interactions, they are decoupled, via the Hubbard-Stratonovich transformation, by introducing an $n \times n$ auxiliary matrix field $Q$.
Through these procedures, the original integral over the large $N \times N$ random systems $H$ is reduced to that over the small $n \times n$ field $Q$.
In the large-$N$ limit $N \to \infty$, the $Q$ integral is dominated by the saddle points, yielding Eq.~\eqref{eq: NLSM}
The possible saddle-point manifold is generally fixed by symmetry of $H$, thereby determining the corresponding universality class.

Notably, the replica formula in Eq.~\eqref{eq: replica-Hermitian} is inapplicable to non-Hermitian systems $H$ with complex spectra $E \in \mathbb{C}$.
Indeed, the density of the complex spectrum is obtained as $\rho \left( E \right) = \pi^{-1} \partial G/\partial E^{*}$, where the identity of the two-dimensional delta function, $\partial E^{-1}/\partial E^{*} = \pi \delta^{(2)} \left( E \right)$, is now relevant.
Accordingly, the replica partition function must be replaced by
\begin{equation}
    G \left( E \right) = \frac{1}{N} \lim_{n\to 0} \frac{1}{n} \frac{\partial}{\partial E} Z_n \left( E \right), \quad Z_n \left( E \right) \coloneqq \braket{[ \det \left( E-H \right) ( E^{*} - H^{\dag} ) ]^n}.
        \label{eq: replica-non-Hermitian}
\end{equation}
As in the Hermitian case, nonlinear sigma models are developed for this modified replica partition function $Z_n \left( E \right)$.
Importantly, this is equivalent to the replica partition function of Hermitized systems~\cite{Girko-85, Feinberg-97},
\begin{equation}
    \tilde{H} \left( E\right) \coloneqq \begin{pmatrix}
        0 & H-E \\
        H^{\dag} - E^{*} & 0
    \end{pmatrix},
        \label{eq: Hermitization}
\end{equation}
which respects additional CS by construction:
\begin{equation}
    \Gamma \tilde{H} \left( E\right) \Gamma^{-1} = - \tilde{H} \left( E\right), \quad \Gamma \coloneqq \begin{pmatrix}
        1 & 0 \\
        0 & -1
    \end{pmatrix}.
\end{equation}
As a consequence of this extra CS, the relevant target manifolds differ from those in the Hermitian cases;
for comparison, see the Hermitian (Table~\ref{tab: Hermitian})
and the non-Hermitian (Tables~\ref{tab: complex AZ}, \ref{tab: real AZ}, \ref{tab: real AZ-dag}, and \ref{tab: real AZ + SLS}) cases.
For example, in symmetry classes A, AI$^{\dag}$, and AII$^{\dag}$,
the functional integral representation of $Z_n(E)$ enjoys
$G\times G$ symmetry due to CS. Here,  $G=\mathrm{U}\left(n\right), \mathrm{Sp}\left(n\right)$, and 
$\mathrm{O}\left(n\right)$ for classes A, AI$^{\dag}$, and AII$^{\dag}$,
respectively.
This symmetry is spontaneously broken down to the diagonal subgroup, $G\times G\to G$, resulting in the target manifold
$(G\times G)/G \cong G$. 
%locking the two chiral sectors  in Eq.\ \eqref{eq: Hermitization} together.
This is an example of strong-to-weak symmetry breaking
\cite{Lee_2023,
Ma_2025,
sala2024spontaneousstrongsymmetrybreaking,
Lessa_2025}.
%As a consequence of this extra CS, the relevant target manifolds differ;
%Compare the Hermitian case in Table~\ref{tab: Hermitian} with the non-Hermitian case in Tables~\ref{tab: complex AZ}, \ref{tab: real AZ}, \ref{tab: real AZ-dag}, and \ref{tab: real AZ + SLS}.
These different target manifolds, in turn, give rise to the distinct universality classes of spectral statistics, as well as Anderson transitions (see Sec.~\ref{sec: Anderson} for details).
In passing, the replica partition function $Z_n \left( E \right)$, or equivalently characteristic polynomial, is of independent mathematical interest (see, for example, Refs.~\cite{Brezin-00, Keating-00}).

Within the framework of nonlinear sigma models, the bulk spectral statistics in classes AI$^{\dag}$ and AII$^{\dag}$~\cite{Kulkarni-25}, and the hard-edge statistics in class D~\cite{Chen-25} were obtained analytically.
These calculations qualitatively capture the deviations from the Ginibre ensembles discussed in Sec.~\ref{sec: NH RMT symmetry}, albeit with quantitative discrepancies stemming from subtleties of the replica method.
In this regard, it is worth noting that the supersymmetry~\cite{Efetov-textbook} and Keldysh~\cite{Kamenev-textbook} formulations offer equally powerful routes to nonlinear sigma models.
Moreover, nonlinear sigma models also describe the nonunitary dynamics of free fermions subject to quantum measurements~\cite{Jian-22, Jian-23, Fava-23, Poboiko-23};
the connection to non-Hermitian disordered systems merits further exploration.
As another application, nonlinear sigma models capture semiclassical periodic orbits in closed quantum systems~\cite{Haake-textbook}.
Extensions to non-Hermitian disordered settings are expected to be relevant for characterizing the chaotic behavior in open quantum systems, which is left for future studies (see also Sec.~\ref{sec: dissipative quantum chaos}).

%%%%%%%%%%%%%%%%%%%%%%%%
\section{DISSIPATIVE QUANTUM CHAOS}
    \label{sec: dissipative quantum chaos}

One of the central physical applications of random matrix theory is the characterization of quantum chaos~\cite{Haake-textbook}.
Bohigas, Giannoni, and Schmit conjectured that a closed quantum system exhibits the real-spectral statistics of Hermitian random matrices if their classical limit is chaotic~\cite{BGS-84}.
Meanwhile, Berry and Tabor conjectured that if the classical limit is integrable, the corresponding quantum spectrum instead displays the Poisson statistics~\cite{Berry-Tabor-77}.
This quantum-classical correspondence has been verified in numerous numerical studies across a wide range of models and underlies the foundation of quantum statistical mechanics~\cite{Huse-review, Rigol-review}.

While the quantum-classical correspondence has been explored primarily in closed systems, it is equally significant in open systems, where classical chaos is likewise ubiquitous.
For example, the Lorenz equation~\cite{Lorenz-63}, prototype of chaotic dynamics, constitutes a set of nonlinear differential equations for dissipative systems.
Remarkably, Grobe, Haake, and Sommers extended the Bohigas-Giannoni-Schmit conjecture by proposing that open quantum chaotic systems exhibit the complex-spectral statistics of non-Hermitian random matrices, while their integrable counterparts display the complex Poisson statistics~\cite{Grobe-88}.
Specifically, they substantiated this conjecture by numerically demonstrating that the level-spacing distribution of a periodically-kicked top with damping in the classically chaotic (regular) regime follows the Ginibre distribution in Eq.~\eqref{eq: Ginibre level-spacing distribution} (Poisson distribution in Eq.~\eqref{eq: Poisson}).
Recently, the Grobe-Haake-Sommers conjecture has been revisited and extensively investigated across various open quantum many-body systems~\cite{Hamazaki-19, Hamazaki-20, Akemann-19, Sa-20}.

\begin{figure}[t]
\centering
\includegraphics[width=1.0\linewidth]{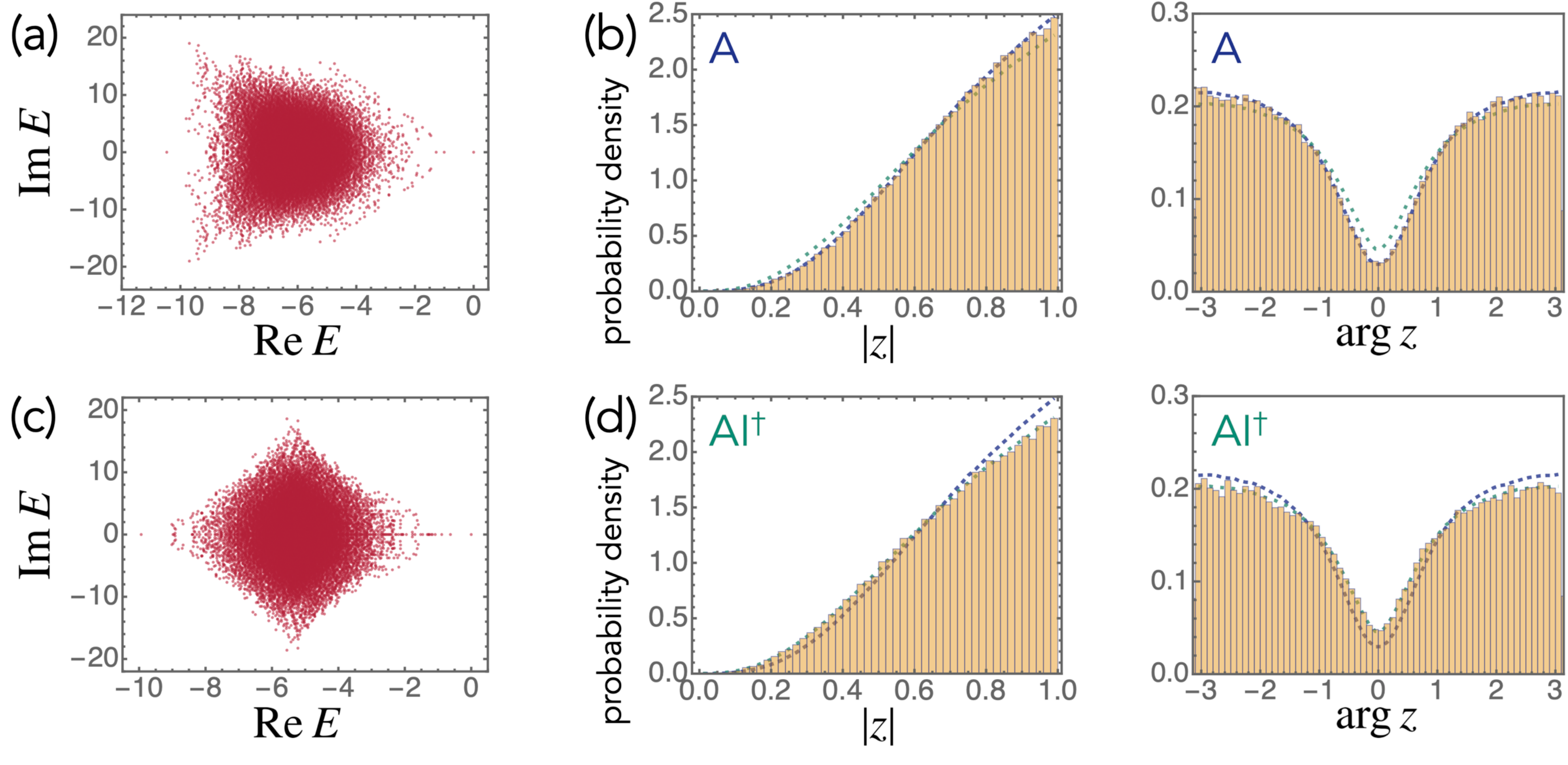}
\caption{Complex-spectral statistics of the (a, b)~damped and (c, d)~dephased Ising models [$L=7$, $J=1.0$, $h_x = -1.05$, $h_z = 0.2$, $\varepsilon = 0.2$; (a, b)~$\gamma = 1.75$ and (c, d)~$\gamma = 0.75$].
(a, c)~Complex spectrum for a single realization.
(b, d)~Probability density of complex level-spacing ratios $z$ for their modulus $\left| z \right|$ and argument $\arg z$.
The numerical data are taken from complex eigenvalues away from the real axis and spectral edges, and averaged over $20$ disorder realizations.
Blue and green dashed curves: 
non-Hermitian random matrices in the Gaussian ensembles for
classes A and AI$^{\dag}$.}
    \label{fig: Lindblad}
\end{figure}

As a representative example, we consider open quantum many-body systems described by the Lindblad master equation~\cite{Nielsen-textbook, Breuer-textbook, Rivas-textbook},
\begin{equation}
    \frac{d\rho}{dt} = \mathcal{L} \left( \rho \right) \coloneqq -\ii \left[ H, \rho \right] + \sum_{n} \left[ L_n \rho L_n^{\dag} - \frac{1}{2} \{ L_n^{\dag} L_n, \rho \} \right],
\end{equation}
where $H$ is a Hermitian Hamiltonian generating the coherent evolution, and $L_n$'s are dissipative jump operators encoding the coupling to the surrounding environment.
We take $H$ to be a nonintegrable quantum Ising model with both longitudinal and transverse fields,
\begin{equation}
    H = - J \sum_{n=1}^{L-1} \left( 1+\varepsilon_n \right) \sigma_n^z \sigma_{n+1}^z - \sum_{n=1}^{L} \left( h_x \sigma_n^x + h_z \sigma_n^z \right),
\end{equation}
with the open boundary conditions.
Here, $\varepsilon_n$ is independently sampled from $\left[ -\varepsilon, \varepsilon \right]$ ($\varepsilon \geq 0$) for each bond so as to lift accidental symmetry.
For the dissipators, we choose 
\begin{equation}
    L_n = \begin{cases}
        \sqrt{\gamma} \sigma_n^{-} & \left( \text{damping} \right); \\
        \sqrt{\gamma} \sigma_n^z & \left( \text{dephasing} \right).
    \end{cases}
\end{equation}

To access the spectrum of the Lindbladian superoperator $\mathcal{L}$, 
we employ the standard vectorization map, sending the density operator $\rho = \sum_{i, j} \rho_{ij} \ket{i} \bra{j}$ to a pure state $\ket{\rho} = \sum_{i, j} \rho_{ij} \ket{i} \ket{j}$ in the double Hilbert space. 
Under this vectorization procedure $\rho \mapsto \ket{\rho}$, the Lindblad equation becomes $\left( d/dt \right) \ket{\rho} = \mathcal{L} \ket{\rho}$, where $\mathcal{L}$ acts as a generally non-Hermitian operator on the double Hilbert space:
\begin{equation}
    \mathcal{L} = -\ii \left( H \otimes I - I \otimes H^{*} \right) + \sum_{n} \left[ L_n \otimes L_n^{*} - \frac{1}{2} \left( L_n^{\dag} L_n \otimes I + I \otimes L_n^T L_n^*\right) \right].
\end{equation}
Arbitrary Lindbladians $\mathcal{L}$ must preserve Hermiticity of density operators, leading to invariance under modular conjugation, $\mathcal{J} \mathcal{L} \mathcal{J}^{-1} = \mathcal{L}$, with an antiunitary operator $\mathcal{J}$ exchanging the bra and ket sectors.
Within the 38-fold classification of non-Hermitian operators, modular-conjugation symmetry is identified with TRS in Eq.~\eqref{eq: TRS} with $\mathcal{J}^2 = +1$.
Accordingly, the damped Ising model is assigned to class AI.
On the other hand, the dephased Ising model additionally respects TRS$^{\dag}$ in Eq.~\eqref{eq: TRS-dag}, $\mathcal{L}^T = \mathcal{L}$, and thus belongs to class BDI$^{\dag}$.
More comprehensive classification has been developed for single-particle~\cite{Lieu-20} and many-body~\cite{Sa-23, Kawabata-23} Lindbladians.

In Fig.~\ref{fig: Lindblad}, we show the distributions of level-spacing ratios in Eq.~\eqref{eq: level-spacing ratio} for the damped and dephased Ising models.
In these numerical simulations, we sample complex eigenvalues sufficiently far from the real axis and spectral edges to extract the bulk spectral statistics.
The obtained data agree with the random-matrix statistics in classes A and AI$^{\dag}$.
Furthermore, modular-conjugation symmetry changes the spectral correlations around the real axis, leading to the suppression of the spectral density as in Eq.~\eqref{eq: DoS - real axis}
Notably, the Lindbladians $\mathcal{L}$ are represented by sparse matrices because of locality, in stark contrast to fully-connected random matrices. 
Nevertheless, the complex-spectral statistics of $\mathcal{L}$ follow the random-matrix statistics,
providing a defining feature of nonintegrability in open quantum systems.
While we have focused on open quantum spin systems, the complex-spectral statistics have also been explored in open quantum fermionic systems, such as the dissipative Sachdev-Ye-Kitaev models~\cite{GarciaGarcia-22PRX, Kawabata-23}.
Moreover, a recent experimental simulation of digital quantum circuits on a high-fidelity superconducting quantum processor has measured level-spacing-ratio distributions and presented clear signatures of level repulsion~\cite{Wold-25}.

Remarkably, the breakdown of the Grobe-Haake-Sommers conjecture has recently been reported~\cite{Villasenor-24, Mondal-25, Griffith-25, Villasenor-25}.
In particular, the random-matrix bulk spectral statistics and the chaotic dynamics in the semiclassical limit can coexist in the open Dicke model~\cite{Villasenor-24}.
A similar inconsistency has also been identified in the original damped periodically-kicked top studied by Grobe, Haake, and Sommers~\cite{Villasenor-25}.
Indeed, the diagnostics discussed above are defined purely in terms of distances in the complex plane, even though the real and imaginary parts of the complex spectrum play qualitatively distinct roles in the open quantum dynamics.
These observations call for further investigation into the dissipative quantum chaos and the quantum-classical correspondence in open systems.
In this respect, it is also notable that while we have focused on complex-spectral statistics, scrambling dynamics in open quantum systems has likewise been investigated~\cite{Syzranov-18, Zhang-19, Touil-21, Zanardi-21, Andreadakis-23, Bhattacharya-22, Liu-23, Schuster-23, Bhattacharya-23, Srivatsa-24}.

%%%%%%%%%%%%%%%%%%%%%%%%
\section{DISORDER-INDUCED CRITICALITY}
    \label{sec: Anderson}

One of the most striking disorder-induced phenomena is Anderson localization and transition~\cite{Anderson-58, Abrahams-79}, which profoundly influence a wide range of condensed matter systems~\cite{Evers-review}.
We review Anderson transitions in non-Hermitian disordered systems, from the Hatano-Nelson model in Sec.~\ref{sec: Hatano-Nelson} to more generic models in Sec.~\ref{sec: Anderson-generic}.

\begin{figure}[t]
\centering
\includegraphics[width=1.0\linewidth]{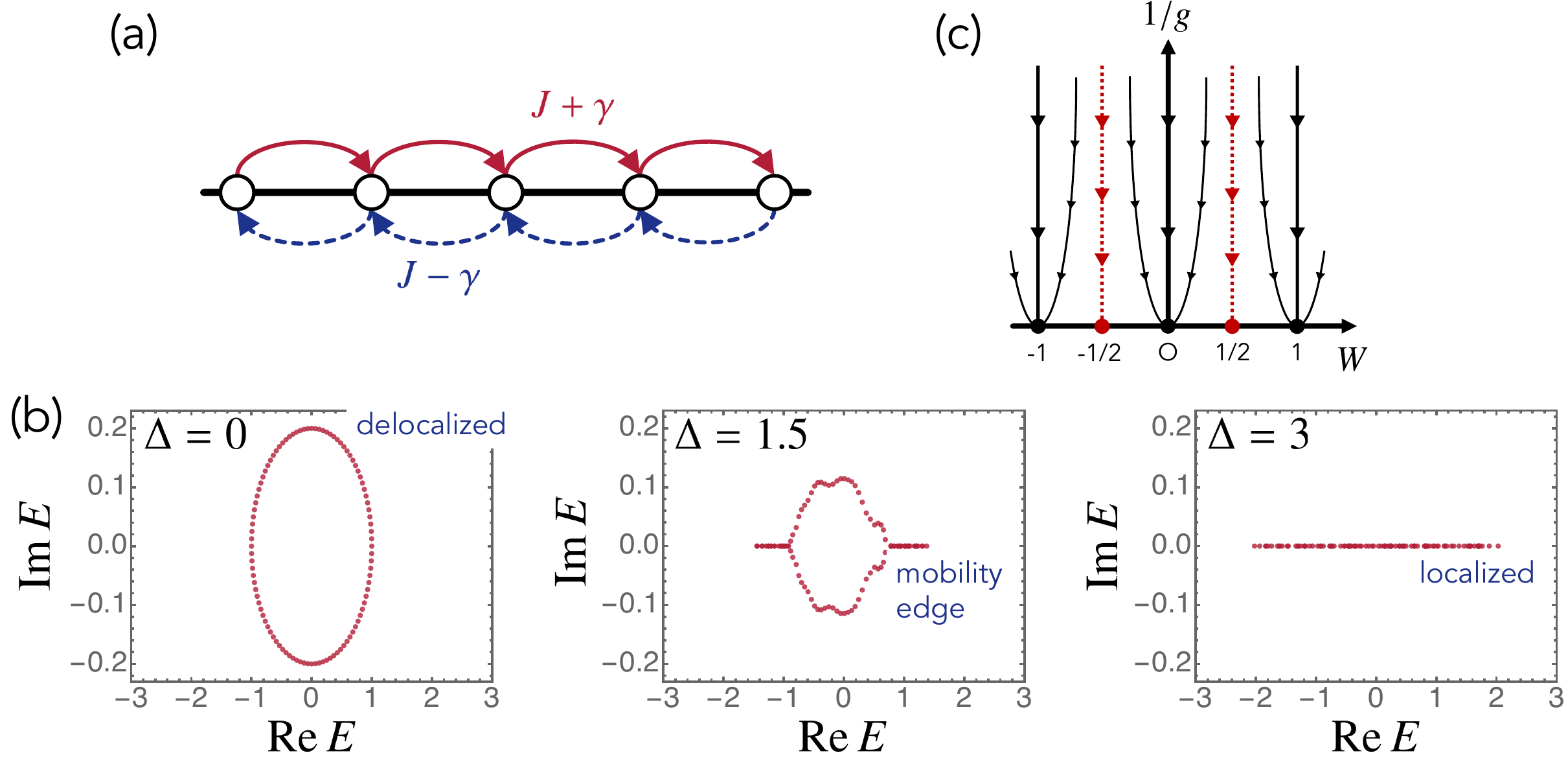}
\caption{Hatano-Nelson model.
(a)~Schematic of the nonreciprocal disordered chain.
(b)~Complex spectra for single realizations of disordered potentials with strength $\Delta = 0, 1.5, 3$ ($J=1.0$, $\gamma = 0.2$).
(c)~Renormalization-group flow.
Two-parameter scaling in terms of the normal coupling parameter $1/g$ and the topological term $W$.
In addition to the fixed points with $1/g = 0$, $W\in \mathbb{Z}$ (black dots), the fixed points with $1/g = 0$, $W \in \mathbb{Z} + 1/2$ (red dots) appear.}
    \label{fig: Hatano-Nelson}
\end{figure}

%%%%%%%%%%%%
\subsection{Hatano-Nelson model}
    \label{sec: Hatano-Nelson}

A prototypical model exhibiting an Anderson transition unique to non-Hermitian disordered systems was introduced by Hatano and Nelson~\cite{Hatano-Nelson-96, Hatano-Nelson-97}.
It describes a nonreciprocal disordered chain with the Hamiltonian [Fig.~\ref{fig: Hatano-Nelson}\,(a)],
\begin{equation}
    H = \sum_{n=1}^{L} \left( - \frac{J+\gamma}{2} c_{n+1}^{\dag} c_n - \frac{J-\gamma}{2} c_{n}^{\dag} c_{n+1} + m_n c_n^{\dag} c_n \right),
\end{equation}
where $c_n$ ($c_n^{\dag}$) annihilates (creates) a particle at site $n$, 
$J + \gamma$ ($J - \gamma$) denotes the hopping amplitude from the left to the right (from the right to the left), 
and $m_n$ is the random onsite potential. 
Here, $m_n$'s are drawn independently from a uniform distribution on $\left[ -\Delta/2, \Delta/2 \right]$ ($\Delta \geq 0$), and the periodic boundary conditions are imposed (i.e., $c_{L+1} = c_{1}$, $c_{L+1}^{\dag} = c_1^{\dag}$).
In its original context, the Hatano-Nelson model describes a depinning transition of type-II superconductors, where the non-Hermitian term $\gamma$ arises from a tilted external magnetic field~\cite{Hatano-Nelson-96, Hatano-Nelson-97}.
It also represents a non-Hermitian extension of the one-dimensional Anderson model~\cite{Anderson-58}.
More recently, similar nonreciprocal models were realized across various open classical experiments of mechanical systems~\cite{Ghatak-19-skin-exp}, electric circuits~\cite{Helbig-19-skin-exp}, photonic systems~\cite{Weidemann-20-skin-exp}, and active matter~\cite{Palacios-21}, as well as open quantum ones of single photons~\cite{Xiao-19-skin-exp}, ultracold atoms~\cite{Liang-22, Zhao-25}, and digital processors~\cite{Shen-25}.

In the absence of disorder, eigenstates are delocalized plane waves, and the corresponding eigenenergies trace a loop in the complex energy plane [Fig.~\ref{fig: Hatano-Nelson}\,(b)]:
\begin{equation}
    E \left( k \right) = J \cos k + \ii \gamma \sin k \quad \left( 0 \leq k < 2\pi \right).
        \label{eq: Hatano-Nelson}
\end{equation}
Upon introducing disorder, some eigenstates remain delocalized, while others undergo Anderson localization,
between which critical mobility edges emerge.
Accordingly, the spectral loop shrinks gradually.
As we further increase the disorder strength, all the eigenstates become localized, and the spectrum becomes entirely real.
Notably, owing to TRS in Eq.~\eqref{eq: TRS}, the Anderson transition in the Hatano-Nelson model coincides with the spectral transition:
delocalized eigenstates carry complex eigenenergies whereas localized eigenstates real eigenenergies.

A salient distinction from the Hermitian regime is the emergence of delocalized states and Anderson transitions even in one dimension.
For generic Hermitian systems in one dimension, even infinitesimal disorder eliminates delocalized states and localizes all eigenstates, generally guaranteed by the one-parameter scaling hypothesis~\cite{Abrahams-79}.
The presence of delocalized states in the Hatano-Nelson model therefore signals the breakdown of this renowned hypothesis and, indeed, arises from topology intrinsic to non-Hermitian systems~\cite{Gong-18, KSUS-19}.
Specifically, the loop structure of the complex spectrum [see Eq.~\eqref{eq: Hatano-Nelson} and Fig.~\ref{fig: Hatano-Nelson}\,(b)] enables one to define the integer-valued winding number $W$ as
\begin{equation}
    W \left( E \right) \coloneqq \oint_0^{2\pi} \frac{d\phi}{2\pi\ii} \frac{d}{d\phi} \log \det \left[ H \left( \phi \right) - E \right] \in \mathbb{Z},
        \label{eq: winding}
\end{equation}
where $H \left( \phi \right)$ is the Hatano-Nelson Hamiltonian threaded by a magnetic flux.
This complex-spectral winding number is identically zero in Hermitian systems and thus serves as the simplest topological invariant intrinsic to non-Hermitian systems.
It remains quantized unless the complex spectrum collapses, i.e., point gap~\cite{Gong-18, KSUS-19}, and thereby protects the delocalization of eigenstates, analogous to the Thouless criterion~\cite{Thouless-review}.

In terms of nonlinear sigma models in Sec.~\ref{sec: NLSM}, the complex-spectral winding number in Eq.~\eqref{eq: winding} manifests as an additional topological term~\cite{Chen-25}:
\begin{equation}
    S [Q] = \int dx \left\{\frac{1}{2g}
   \mathrm{tr} \left[ ( \partial_x Q^{\dag} ) ( \partial_x Q ) \right] + W \mathrm{tr} \left[ Q^{\dag} \partial_x Q \right] \right\}, \quad Q \in \mathrm{U} \left( n \right).
\end{equation}
The emergence of this topological term clearly indicates the breakdown of the one-parameter scaling and instead leads to the two-parameter scaling in Fig.~\ref{fig: Hatano-Nelson}\,(c), reminiscent of the quantum Hall transition~\cite{QHE-textbook, Huckestein-review, Evers-review}.
%~\cite{Khmel'nitskii-83, Pruisken-84, QHE-textbook, Huckestein-review, Evers-review}.
Beyond the Hatano-Nelson model, possible topological terms of nonlinear sigma models, corresponding to point-gap topology~\cite{KSUS-19}, are classified in Tables~\ref{tab: complex AZ}, \ref{tab: real AZ}, \ref{tab: real AZ-dag}, and \ref{tab: real AZ + SLS}~\cite{Chen-25}.
It is also notable that the nontrivial point-gap topology, as in Eq.~\eqref{eq: winding}, underlies the extraordinary sensitivity to boundary conditions, known as the non-Hermitian skin effect~\cite{Bergholtz-review, Okuma-Sato-review}.
%~\cite{Lee-16, YW-18-SSH, Kunst-18, Lee-Thomale-19, Yokomizo-19, Zhang-20, OKSS-20}.

%%%%% symmetry & spectra %%%%%
\begin{table}[t]
    \tabcolsep0pt
	\centering
	\caption{Critical exponents of Anderson transitions in three-dimensional Hermitian and non-Hermitian systems within the Wigner-Dyson class.
    The quoted uncertainties correspond to 95\% confidence intervals for ${}^{*}$ and double standard deviations for ${}^{**}$.}
	\label{tab: Anderson}
     \begin{tabular}{ccc} \hline \hline
    ~~Class~~ & ~~Hermitian~~ & ~~Non-Hermitian \\ \hline
    A & $1.43 \pm 0.04^{*}$~\cite{Slevin-97, Slevin-16} & $1.00 \pm 0.04^{**}$~\cite{Luo-21B} \\ \hline
    AI & \multirow{2}{*}{$1.57 \pm 0.03^{*}$~\cite{Slevin-97, Slevin-18}} & $0.93 \pm 0.14^{*}$~\cite{Luo-22R} \\ 
    AI$^{\dag}$ & & $1.19 \pm 0.01^{**}$~\cite{Luo-21B} \\ \hline
    AII & \multirow{2}{*}{~~~$1.37 \pm 0.02^{*}$~\cite{Kawarabayashi-96, Asada-05}~~~} & $0.875 \pm 0.004^{*}$~\cite{Luo-22R} \\ 
    ~~~AII$^{\dag}$~~~ & & ~~~$0.903 \pm 0.007^{*}$~\cite{Luo-22R}~~~ \\ \hline \hline
    \end{tabular}
\end{table}

%%%%%%%%%%%%
\subsection{Higher dimensions}
    \label{sec: Anderson-generic}

Beyond the Hatano-Nelson model, Anderson transitions in generic non-Hermitian disordered systems were investigated extensively in recent years~\cite{Longhi-15, Gong-18, Longhi-19, Zeng-20, Tzortzakakis-20, Huang-20, KR-21, Claes-21, Luo-21L, Luo-21B, Luo-22R, Liu-Fulga-21, Kim-21, Moustaj-22, DeTomasi-22, DeTomasi-23, Nakai-24, Ghosh-23, Thompson-24, Chen-25, Li-25}.
The influence of many-body interactions was also studied~\cite{Hamazaki-19}.
In general, the correlation or localization length algebraically diverges around the transition, $\xi \propto \left| r - r_{\rm c} \right|^{-\nu}$, with a critical exponent $\nu > 0$.
Here, $r$ denotes a relevant parameter, such as the disorder strength, and $r_{\rm c}$ is its critical value.
In Hermitian disordered systems, these critical exponents depend not on microscopic details but solely on universal ingredients such as symmetry, dimensions, and topology.
Until recently, however, it has remained unclear whether non-Hermiticity affects the universal critical exponents within the same symmetry classes.

Based on the transfer-matrix method, several recent works~\cite{Luo-21L, Luo-21B, Luo-22R} numerically obtained such universal critical exponents in two and three dimensions, some of which are summarized in Table~\ref{tab: Anderson}.
Importantly, non-Hermiticity does change the critical exponents even within the same symmetry classes, signaling the universality classes distinct from their Hermitian counterparts.
Moreover, the difference between TRS in Eq.~\eqref{eq: TRS} and TRS$^{\dag}$ in Eq.~\eqref{eq: TRS-dag}, which coincide in Hermitian systems, manifests itself through the different critical exponents.
Thus, alongside the random-matrix spectral statistics (see Sec.~\ref{sec: NH RMT symmetry}), the universality classes of Anderson transitions in non-Hermitian disordered systems are likewise classified by the 38-fold way in Sec.~\ref{sec: symmetry}.
It is also notable that the criticality of non-Hermitian disordered systems is identical with that of Hermitized ones in Eq.~\eqref{eq: Hermitization}~\cite{Luo-22R}, 
consistent with the effective field theory description in Sec.~\ref{sec: NLSM}.

%%%%%%%%%%%%%%%%%%%%%%%%
\section{DISCUSSION}
    \label{sec: discussion}

In this review, we have discussed the rich physics and mathematics arising from the interplay between non-Hermiticity and disorder.
In mathematics, research has hitherto concentrated largely on the classic Ginibre ensembles reviewed in Sec.~\ref{sec: Ginibre}, and many analytical properties of generic non-Hermitian random matrices across the 38-fold way in Sec.~\ref{sec: symmetry}, including the most fundamental cases in classes AI$^{\dag}$ and AII$^{\dag}$ (see Sec.~\ref{sec: NH RMT symmetry}), remain outstanding open problems.
In physics, in parallel with the recent rapid advances in quantum technology, developing a deeper understanding of nonequilibrium and/or dissipative quantum dynamics is becoming increasingly important.
Since such quantum dynamics is inherently nonunitary and stochastic, non-Hermitian random matrix theory should likewise play a key role as a framework for modeling complicated open quantum systems.
As discussed in Sec.~\ref{sec: dissipative quantum chaos}, while the complex-spectral statistics provide useful diagnostics of integrability, it remains an outstanding challenge in statistical physics to precisely formulate chaos in open quantum systems.
Nonlinear sigma models in Sec.~\ref{sec: NLSM} may prove instrumental in this context, given their significant role in capturing semiclassical periodic orbits in closed quantum systems.
In high energy physics, the 10-fold way for Hermitian random matrices has counterparts in Jackiw-Teitelboim gravity~\cite{Saad-Shenker-Stanford-19, Stanford-Witten-19}.
It remains an open question whether the 38-fold way for non-Hermitian random matrices likewise admits an analog in gravity theory.
In this respect, it is worth noting that non-Hermiticity has also begun to play a role in recent developments in holography~\cite{Takayanagi-25}.
Finally, while Hermiticity is a fundamental prerequisite for quantum mechanics, no such constraint applies beyond physics.
Rather, statistical properties of non-Hermitian disordered systems should be relevant across the broader natural sciences, for example in neural networks.
We hope that the developments discussed in this review will stimulate further applications throughout a wide range of scientific disciplines.

%%%%%%%%%%%% Disclosure %%%%%%%%%%%%
\section*{DISCLOSURE STATEMENT}
%If the authors have noting to disclose, the following statement will be used: 
The authors are not aware of any affiliations, memberships, funding, or financial holdings that might be perceived as affecting the objectivity of this review. 

%%%%%%%%%%%% Acknowledgements %%%%%%%%%%%%
\section*{ACKNOWLEDGMENTS}
We thank Ze Chen, Giorgio Cipolloni, Ryusuke Hamazaki, Jonah Kudler-Flam, Anish Kulkarni, Naoto Kura, Jiachen Li, Xunlong Luo, Tokiro Numasawa, Tomi Ohtsuki, Keiji Saito, Masatoshi Sato, Ryuichi Shindou, Ken Shiozaki, Masahito Ueda, and Zhenyu Xiao for collaborations and/or discussions.
K.K. is supported by MEXT KAKENHI Grant-in-Aid for Transformative Research Areas A ``Extreme Universe" No.~JP24H00945.
S.R. is supported by the National Science Foundation under Award No.\ DMR-2409412.

\bibliographystyle{ar-style4}
\bibliography{NH}

\end{document}